\documentclass[aps,pra,10pt]{revtex4-1}

\usepackage{amssymb}
\usepackage{amsfonts}
\usepackage{bm}
\usepackage{afterpage}
\usepackage{graphicx}
\usepackage{color}
\usepackage{epstopdf}
\usepackage{amsmath}

\def\beq{\begin{equation}}
\def\eeq{\end{equation}}
\def\nn{\nonumber}
\def\om{\omega}
\def\a{\alpha}
\def\b{\beta}
\def\g{\gamma}
\def\th{\theta}
\def\eps{\epsilon}

\def\s{\sigma}
\def\D{\Delta}

\def\ad{a^\dagger}
\def\rd{{\rm{d}}}

\def\p{\phi}
\def\tp{\tilde{\p}}
\def\dd{{\frac{\rd}{\rd z}}}

\def\ra{\rightarrow}

\def\Cc{\mathbb{C}}
\def\Zz{\mathbb{Z}}
\def\Rr{\mathbb{R}}
\def\Zpl{\rm{I\!N}}

\def\bp{\bar{\phi}}

\def\bE{\bar{E}}
\def\bD{\bar{\D}}

\def\B{{\cal B}}

\def\sqe{\sqrt{\eta}}
\def\tg{\tilde{g}}
\def\tE{\tilde{E}}
\def\tD{\tilde{\D}}
\def\Re{{\rm Re}}
\def\Im{{\rm Im}}

\def\La{\Lambda}
\def\ch{\textrm{ch}}
\def\sh{\textrm{sh}}
\def\tx{\tilde{q}}
\def\ket#1{\lvert\nobreak#1\nobreak\rangle}

\def\Dp{\Delta^{\prime}}
\def\Ep{E^{\prime}}

\begin{document}

  \title{The spectral continuum in the Rabi-Stark model}

 \author{Daniel Braak$^{1,*}$, Lei Cong$^{2,3,**}$, Hans-Peter Eckle$^{4}$, Henrik Johannesson$^{5}$ 
 and Elinor K. Twyeffort$^{6}$
 }

\affiliation{
$^{1}$Institute of Physics, University of Augsburg, 86135 Augsburg, Germany\\
$^{2}$Helmholtz-Institut, GSI Helmholtzzentrum f\"ur Schwerionenforschung, 55128 Mainz, Germany\\
$^{3}$Johannes Gutenberg University, 55128 Mainz, Germany\\
$^{4}$Humboldt Center, University of Ulm, 89081 Ulm, Germany\\
$^{5}$Department of Physics, University of Gothenburg, SE 412 96 Gothenburg, Sweden\\
$^{6}$School of Physics and Astronomy, University of Southampton, Highfield, Southampton, SO17 1B, United Kingdom --
E.K. Twyeffort has previously published as E.K. Irish and E.K. Twyeffort Irish
}

  \begin{abstract}
The Rabi-Stark model is a non-linear generalization of the quantum Rabi model including the dynamical Stark shift as a tunable term, which can be realized via quantum simulation on a cavity QED platform. When the Stark coupling becomes equal to the mode frequency, the spectrum changes drastically, a transition usually termed ``spectral collapse" because numerical studies indicate an infinitely degenerate ground state. We show that the spectrum extends continuously from a threshold value up to infinity. A set of normalizable states are embedded in the continuum which furnishes an unexpected analogy to the atomic Stark effect. Bound states and continuum can be obtained analytically through two equally justified, but different confluence processes of the associated differential equation in Bargmann space. Moreover, these results are obtained independently using a method based on adiabatic elimination of the spin degree of freedom and corroborated through large-scale numerical checks.

  \end{abstract}


  \maketitle

\section{Introduction}\label{sec-intro}

The Rabi model enjoys a long and distinguished career.
Originally introduced by Rabi \cite{Rabi1936}
in 1936 as a semi-classical model to describe the interaction of quantum
mechanical spins with a classical electromagnetic field, its fully
quantum mechanical version was formulated by Jaynes and Cummings
\cite{JC} in 1963.
This model, the quantum Rabi model (QRM), captures the interaction 
of matter (Fermionic discrete degrees of freedom) and light 
(Bosonic continuous degrees of freedom) in a minimalist way
by assuming that a matter qubit, representing a two-level atom, couples
to a single-mode Bosonic field.
Jaynes and Cummings simplified the QRM using the so-called
rotating-wave approximation (RWA), valid for small matter-field interaction, 
to obtain a model now bearing the authors' name,
the quantum Jaynes-Cummings model, which they then solved
by elementary means \cite{JC}.

The rotating-wave approximation enhances the $\Zz_2$-symmetry of the QRM to a continuous $U(1)$-symmetry 
which renders it solvable 
in closed form by dividing the Hilbert space into infinitely many two-dimensional invariant subspaces. 
A similar symmetry enhancement, adapted to the case of strong coupling and called the generalized 
rotating wave approximation (GRWA), 
led again to closed form expressions for eigenenergies and eigenstates, overcoming the failure of the 
RWA for strong coupling \cite{irish2007}. 
It was shown later that the original $\mathbb{Z}_2$-symmetry of the QRM suffices to 
obtain a closed formula for the spectral determinant of the Hamiltonian, the so-called $G$-function \cite{braak2011}.
This symmetry divides the Hilbert space into just two invariant
subspaces, each infinite dimensional, such that level crossings between different
subspaces become possible. However, these exact crossings are far less numerous than in the 
approximations with enhanced symmetry, RWA and GRWA.
The solution of the QRM given in \cite{braak2011} has been reproduced using Bogoliubov operators
\cite{chen2012} by employing a heuristic argument for the implementation of the spectral condition 
(normalizability of the eigenfunctions).

While the quantum Jaynes-Cummings model was sufficient to
interpret experiments for a long time, recent experimental 
developments have been able
to drive the matter-field interaction strength into regimes $-$
the strong, ultra-strong \cite{niemczyk2010} and deep strong 
 \cite{casanova2010}  coupling regimes $-$
where the quantum Jaynes-Cummings model is no longer appropriate,
requiring the full QRM as a replacement. Forays into these
coupling regimes are an active current research area, with examples ranging
from trapped ion systems interacting with a laser field confined to a cavity \cite{haroche2008} 
to superconducting charge qubits in circuit quantum electrodynamics architectures \cite{blais2020,blais2021}.  


A remarkable property of the QRM is that it exhibits
a ``quantum phase transition", showing universal dynamics in the vicinity of criticality \cite{bakemeier2012,ashhab2013,hwang2015}. The appearance of a phase transition in a few-body system comes as a surprise
as such a transition may occur only in the thermodynamic limit of a many-body system, when the particle number $N$ and the volume $V$ diverge such that the density $N/V$ stays finite \cite{sachdev2011}. The thermodynamic limit entails that the level distance between many-body eigenstates goes to zero, a fact which can be mimicked in the QRM 
by the limit $\Delta/\om\ra\infty$, $g/\om\ra\infty$ such that $g/{\sqrt{\om\D}}$ stays finite. The infinite dimensional Hilbert space of the single bosonic mode may replace thus the state space of infinitely many particles, at the cost of introducing experimentally inaccessible parameters. Moreover, this few-body phase transition is of mean-field character \cite{larson2017}.

There are many generalizations of the QRM, inter alia the
two-qubit QRM \cite{peng2012, sun2020}, the two-photon QRM \cite{duan2016,cong2019,rico2020, braak2023}, and the asymmetric QRM \cite{braak2011}. 
The asymmetric version of the model is particularly intriguing and is currently attracting much attention. While the additional driving term breaks the $\Zz_2$-invariance and generically lifts the level crossings of the QRM, the crossings reappear for certain discrete values of the amplitude of the driving term \cite{braak2011,ashhab2020}.
This observation has led to the search for hidden symmetries in the asymmetric QRM and the construction of corresponding operators 
\cite{mangazeev2021,li2021a,Kimoto2021,reyes2021,li2021b,xie2021},
in addition to an analysis of entanglement resonances as a symmetry probe \cite{shi2022}. 

A review of the QRM and several of its generalizations is given in
\cite{xie2017}. Recent reviews of light-matter interaction models and methods
for their solution can be found in \cite{leboite2020} and, with an emphasis
on the QRM, in \cite{larson2021}.
For a recent treatment of the derivation from first principles and an elementary discussion
of the QRM, see \cite{eckle2019}. 

Some ten years ago Grimsmo and Parkins proposed yet another generalization of the QRM \cite{grimsmo2013}
which will be the model further investigated in the present work.
These authors suggested an experimental set-up singling out two hyperfine ground states
of a multilevel atom constituting an effective two-level quantum system
and coupled to a quantized cavity field (single-frequency photons) and two auxiliary laser fields.
The quantum Rabi model realized in this way permits the free and
independent choice of frequencies and coupling constants of the model
such that also the ultrastrong and deep-strong coupling regimes can be
accessed experimentally.
Unless the two laser frequencies are fine-tuned, i.e. for generic values of the
model parameters, an additional, nonlinear, term appears in the
model Hamiltonian  coupling the two-level system and the cavity field.
Such a term is well known as dynamical Stark shift, a quantum
analog of the Bloch-Siegert shift \cite{klimov2009}.
The model is meanwhile known under the name of quantum Rabi-Stark model
(QRSM).
In the dynamical Stark shift, the nonlinear coupling strength is usually dependent
on the other system parameters of the model while in the Grimso-Parkins scheme
also the Stark coupling can be adjusted freely and independently.
The emergence of selective $k$-photon interactions in the strong and ultra-strong regimes 
and an alternative method for the quantum simulation of the QRSM was recently demonstrated \cite{cong2020}.
The latter uses the internal and vibrational degrees of freedom of a trapped ion set-up.

Grimsmo and Parkins \cite{grimsmo2014} discuss the possibility that the QRSM can exhibit a superradiant phase transition in the deep-strong coupling regime 
when the magnitude of the Stark coupling becomes equal to the frequency of the cavity field, i.e., at finite values of the parameters, thus avoiding the unphysical 
parameter regime which led to the phase transition in the QRM \cite{hwang2015}. Spurred by this conjecture, there has been a number of works analyzing the spectral structure 
of the QRSM \cite{maciejewski2014,maciejewski2015,eckle2017,eckle2023,xie2019}. Despite these efforts, a number of intriguing questions remain open, some of which will be 
addressed in this paper. 

Before proceeding, let us summarize what is already known about the spectral structure of the QRSM. As a backdrop
we begin by outlining the basic properties of the
QRM whose
Hamiltonian is
\beq
H_R=\om\ad a +\D\s_z +g\s_x(\ad +a).
\label{hamr}
\eeq
Here 
$\ad$
 and 
$a$
 are Bosonic creation and annihilation operators, respectively,
and $\sigma_z$ and $\sigma_x$ are Pauli spin matrices.
The parameters of the Hamiltonian are the oscillator frequency
$\omega$, the splitting of the two-level system $\Delta>0$, and the Rabi coupling $g>0$. 
The Hamiltonian (\ref{hamr})  possesses a $\mathbb{Z}_2$ or parity symmetry, i.e., the commutator
$[H_R,P]$ vanishes with the parity operator given by
\begin{equation}
P=(-1)^{\ad a}\s_z
\end{equation}
and the Hilbert space ${\mathcal H}$ decomposes into two subspaces 
${\mathcal H}={\mathcal H}_+\oplus{\mathcal H}_-$.
The eigenstates and eigenvalues of this Hamiltonian can thus be labeled by two quantum numbers:
the parity quantum number $p=\pm 1$ and an eigenvalue $n=0,1,2,\ldots$ numbering the energy
eigenvalues from below. The spectral structure of this Hamiltonian \cite{braak2011}  
accordingly shows a positive and a negative parity part where level crossings are
only possible between energy levels of different parities while levels with the same parity
avoid level crossings.
The spectrum of (\ref{hamr}) can also be described as consisting of a regular part to which
almost all eigenvalues belong and an exceptional part formed by the level crossings
of opposite parity levels.

The QRSM with model Hamiltonian\cite{note1}     
\beq
H_{RS}=\om\ad a +\s_z\left(\g\ad a +\D\right)+g\s_x(\ad +a )
\label{hamrs}
\eeq
also enjoys the $\mathbb{Z}_2$ symmetry and, hence, its spectrum can, on the whole, be
described similar to the QRM spectrum. Indeed, we shall show in section \ref{sec-qrm} that the coupled system of ordinary differential equations (ODE's) representing the eigenvalue equation for $H_{RS}$ can be mapped onto the corresponding system of ODE's for $H_R$, albeit with certain modifications which result in qualitative changes which we now describe phenomenologically. 

The first characteristic of the QRSM is that, whereas in the QRM each successive pair of spectral
levels labeled by $\{p=\pm1,n=0,1,2,\ldots\}$ exhibits $n$ level crossings before it eventually
becomes degenerate for Rabi coupling $g \rightarrow \infty$, the QRSM exhibits one more level crossing
before the levels again become pairwise degenerate for $g\rightarrow\infty$.
The additional level crossings occur for \cite{xie2019}
\beq
g_c^{(n)} = \sqrt{\left( n + \frac{\Delta}{\gamma}\right)} \sqrt{(\omega^2-\gamma^2)}
\qquad n=0,1,2,\ldots.
\label{gcn}
\eeq
Note that all $g_c^{(n)}\rightarrow\infty$ as $\gamma\rightarrow0$, reproducing the behaviour of the QRM spectrum. 
The first of the additional crossings, between the ground state and the first excited state, 
occurring at $g_c^{(0)}$, has been interpreted as a first-order quantum phase transition (QPT) \cite{xie2019} which occurs in similar way in the anisotropic QRM \cite{chen2021}.

Keeping the Stark coupling $\gamma$ finite leads to changes in the overall structure of
the spectrum.
One subset of the spectral levels start to move upwards as $\gamma$ increases (we assume throughout the following $\gamma, \D\ge 0$).
These are the levels which, for $g=0$, start at
\beq
E(g=0)=(\om+\gamma)n+\Delta \qquad n=0,1,2,,\ldots.
\eeq
Another subset of levels start to move downward as  $\gamma$ increases and can be identified
as starting from
\beq
E(g=0)=(\om-\gamma)n-\Delta \qquad n=0,1,2,,\ldots.
\eeq
While the former levels, for $\gamma\rightarrow\omega$, 
eventually tend to the spectrum of a harmonic oscillator,
the latter tend, in the same limit, to a massively degenerate spectrum at $E=-\D$.
However, as we shall see in the following, there is a dramatic restructuring of the spectrum  when switching on $g$: When $\gamma\rightarrow\om$ the latter branch of energy levels now fuse into a continuum of states similar to the QRM with non-linear (two-photon) coupling when the coupling reaches a critical value \cite{duan2016,rico2020,braak2023,ng1999,felicetti2015}. 

We shall have occasion to interpret the spectral structure of the QRSM in the
limiting case $\gamma \rightarrow \omega$ with $g > 0$ as spectral levels of bound states embedded in a continuum
of non-normalizable states.
This interpretation is borne out by the results of large scale numerical exact 
diagonalization calculations which reveal the phenomenological picture of the
spectrum outlined above.
Moreover, analytical calculations demonstrate explicitly what the numerical
spectra suggest and, furthermore, in the limit $\gamma \rightarrow \omega$,
prove the existence of discrete parts of the
spectrum corresponding to bound states, and continuous parts of the
spectrum corresponding to extended states.
The bound states are embedded in the continuum and can thus be identified as
``bound states in the continuum" (BICs), a phenomenon which has attracted
considerable experimental and theoretical attention in recent years, see the review
\cite{BICreview} and the following discussion. 

Returning to the original proposal for the QRSM \cite{grimsmo2013}, 
Grimsmo and Parkins noted that an instability develops in the system as the magnitude of the 
Stark coupling $|\gamma|$ grows towards the photon frequency $\omega$, 
causing sharp changes in the photon number and other observables in the deep-strong-coupling regime $g > \omega$. 
It was conjectured that $\gamma = \pm \omega$ define critical points that mark the occurrence of 
quantum phase transitions, with the ground states here becoming massively degenerate. Specifically $-$ 
based on a semiclassical analysis of the reduced density matrix in the presence of cavity dissipation $-$ 
the system was predicted to undergo a transition to a superradiant phase at $\gamma = -\omega$ \cite{grimsmo2013}. 

Independent support for the existence of possible phase transitions at $\gamma = \pm \omega$ was 
provided by Xie {\em et al.}, using a mapping of the QRSM Hamiltonian (\ref{hamrs}) at the critical point 
$\gamma=\omega$ to an effective quantum harmonic oscillator \cite{xie2019}. 
These authors found that part of the discrete energy spectrum collapses onto a massively degenerate level at a critical 
Rabi coupling $g_c = \sqrt{(\om/2-\Delta)\omega}$, exhibiting manifest scaling behavior of the photon number 
indicative of a quantum phase transition \cite{chen2020,note2} Intriguingly, for all $g>0$, still at $\gamma = \pm \omega$, 
there is the possibility that the spectral gap for energies $E$ in the interval
$-\Delta - 2g^2/\omega < E < -\Delta$ may get filled by a continuous band of energies. If so, this interval supports
an uncountable set of non-normalizable (scattering) states. A similar appearance of non-normalizable states in the 
two-photon Rabi model is well known \cite{rico2020,braak2023} and was anticipated also for 
the QRSM by Maciejewski {\em et al.} \cite{maciejewski2015}. However, physical implications $-$ as well 
as a rigorous proof of the very existence of a continuum of states $-$ have remained elusive. 

In this article we revisit the problem, taking off from a set of scale transformations that map 
the Hamiltonian in Equation (\ref{hamrs}) to an effective quantum Rabi model. Similar to the ordinary QRM, 
the eigenvalue equation becomes equivalent to a coupled system of first-order ODE's in 
Bargmann space \cite{braak2011}. The two regular singular points of this system are found to coalesce 
to an irregular singular point at infinity in the limit $\gamma \rightarrow \omega_{-}$. 
By studying the confluence process, with the standard assumption that the variable prefactors in the system 
of ODE's are bounded with respect to the diverging parameters, we can furnish a formal proof that there is a
spectral continuum with non-normalizable states between $- \Delta -2g^2/\omega$ and $-\Delta$. For $E > - \Delta$ 
the spectrum remains discrete with normalizable eigenstates.

Lifting the assumption that the confluence process comes with a certain hierarchy of diverging 
quantities unveils a dramatic scenario: We can now prove the emergence of an unbounded continuous energy band for 
{\em all} energies $E > -\Delta-2g^2/\omega$ in the limit $\gamma \rightarrow \omega$. 
The corresponding non-normalizable (scattering) states are orthogonal to the normalizable 
states of the discrete spectrum for $E > -\Delta$, implying that the latter states get embedded in a spectral continuum. 
It follows that the normalizable states can be considered as {\em bound states in the continuum} 
(BICs) \cite{BICreview,koshelev2021,sadreev2021} at the point $\gamma = \omega$ (with the expectation $-$ yet to be proven $-$ 
that the same scenario plays out also at $\gamma = -\omega$).

The anomalous confinement phenomenon implied by BICs has attracted a lot of interest recently, primarily brought on by
potential applications in photonics \cite{kodigala2017,longhi2021,qin2022,kang2023}.    
Quite generally, the existence of a BIC $-$ be it in the classical or quantum realm $-$ depends 
on some mechanism that confines the state despite it being degenerate with an extended scattering state 
(that would otherwise radiate away the BIC under an infinitesimal perturbation). 
Numerous confinement mechanisms have been proposed over the years $-$ 
from {\em tailored potentials} (oscillating in such a way as to support a BIC, 
as in the pioneering work by von Neumann and Wigner \cite{neumann1929}), to {\em symmetry protection} 
(with no coupling between the BIC and the extended states if belonging to different 
symmetry classes \cite{plotnik2011,zhang2012}) and {\em separation of variables} (with each term of a 
separable Hamiltonian having a bounded eigenstate in addition to a continuous band \cite{jain1975,robnik1986}). 

While these classes of BICs and their origins have been well studied $-$ albeit still challenging to 
realize in a quantum system $-$ the BICs we find in the QRSM are of a different kind. 
The only requirement here is the fine tuning that sets $\gamma = \omega$. One may think that this is 
an instance of a {\em parametric} BIC, yet another class of such states (where a BIC gets confined by 
tuning the radiating channels so that they cancel \cite{hsu2013}). However, different from generic parametric BICs 
known from the literature, only a {\em single} parameter needs tuning in our case, making it potentially more accessible 
in an experiment.

Importantly, the QRSM exhibits precursors of BICs in the neighborhood of $\gamma=\omega$. 
We shall call the corresponding states {\em preBICs} and part of our work will focus on these 
states as revealed by numerics. Their existence and their distinct 
features may add to the experimental accessibility of this novel type of BICs. 

The paper is organized as follows: In section \ref{sec-qrm}, we calculate the spectral determinant ($G$-function) $G_\pm(x)$ of the QRSM in a simplified way by mapping it to an effective QRM. In this way, results obtained in \cite{xie2019} about a novel class of degeneracies and the concomitant first order quantum phase transition are easily explained.
In section \ref{sec-limit}, we study the limit $\g\ra\om_-$.
The pole structure of $G_\pm(x)$ gives information about the average level spacing and thus a first hint to the presence of a continuum at $\g=\om$ for $E>-\D$. There are two possible confluence procedures: the first one is standard and has been employed in \cite{maciejewski2015,xie2019}. The second one is novel and yields the full continuum together with the embedded bound states. In section \ref{sec-num} we give independent corroborating evidence for the full continuum by numerical exact diagonalization of the model in state spaces with very high dimension. Section \ref{sec-slowmode} is devoted to an independent confirmation through the explicit analytical treatment of the QRSM under the assumption that the radiation mode is slow compared to the qubit. Finally, we present our conclusions and outlook to future research directions in section \ref{sec-concl}.  

\section{Effective quantum Rabi model}\label{sec-qrm}
The following analysis uses the representation of the creation operator
$\ad$ as multiplication with the complex variable $z$ and the annihilation operator $a$ as derivative
$\rd/\rd z$. 
These operators are mutually adjoint in the Bargmann space $\B$ of analytic functions \cite{bargmann1961}.
States $\psi(x)\in L^2(\mathbb{R})$, the Hilbert space of square integrable functions on the real line, are 
mapped to analytic functions $\psi(z)$ of $z$. The inner
product is defined as
\beq
\langle\psi|\phi \rangle=\frac{1}{\pi}\int \rd z\rd \bar{z}\overline{\psi(z)}\phi(z)e^{-z\bar{z}}.
\eeq
In the Bargmann representation there are two requirements for a function $\psi(z)$ to be an element of $\B$. 
The first Bargmann condition requires analyticity of $\psi(z)$ in all of $\Cc$, $\psi(z)$ must be entire. 
The second Bargmann condition reads $\langle\psi|\psi\rangle<\infty$. 
Both conditions will play a crucial role in the exact determination of the spectrum of 
$H_{RS}$ given in Equation (\ref{hamrs}).

In the Bargmann space $\B$, the eigenvalue equation for $H_{RS}$ in (\ref{hamrs}),
\begin{equation}
H_{RS}\bm{\phi} = E\bm{\phi}
\end{equation}
gets represented by the coupled system of first-order ODE's \cite{maciejewski2015}
\begin{eqnarray}
  (\om+\g)z\dd\p_1+\D\p_1 +g\left(z+\dd\right)\p_2 &=E\p_1,
  \label{orig1}\\ 
  (\om-\g)z\dd\p_2-\D\p_2 +g\left(z+\dd\right)\p_1 &=E\p_2,
  \label{orig2}
\end{eqnarray}
with $\bm{\phi} = (\phi_1,\phi_2)^T$.
First, we make the following scale transformation,
\beq
\p_1(z)=\tp_1(z), \qquad \p_2(z)=\eta\tp_2(z), \quad \eta=\frac{\om+\g}{\om-\g}.
\label{scalet}
\eeq
This transformation becomes singular at $\om=\g$; we assume for now $\om>\g$ and study 
the limit $\g\ra \om_{-}$ later on. 
The Equations (\ref{orig1},\ref{orig2}) read now $\hat{A}\bm{\tp}=\bm{0}$, with
\beq
\hat{A}=
\left(\!\!\begin{array}{cc}
  1&0\\
  0&1
\end{array}\!\!\right)(\om+\g)z\dd +
\left(\!\!\begin{array}{cc}
\D-E&0\\
  0&-\eta(\D+E)
\end{array}\!\!\right) +
\left(\!\!\begin{array}{cc}
0&\eta\\
  1&0
\end{array}\!\!\right)g\left(z+\dd\right).
\label{matrixeq}
\eeq
We may now diagonalize the term proportional to $z+\rd/\rd z$,
\beq
\hat{U}^{-1}
\left(\!\!\begin{array}{cc}
  0&\eta\\
  1&0
\end{array}\!\!\right)
\hat{U}=
\left(\!\!\begin{array}{cc}
  -\sqrt{\eta}&0\\
  0&\sqrt{\eta}
\end{array}\!\!\right), \qquad
\hat{U}=
\left(\!\!\begin{array}{cc}
  \sqrt{\eta}&\sqrt{\eta}\\
  -1&1
\end{array}\!\!\right).
\eeq
It follows with $\bm{\tp}=\hat{U}\bm{\bp}$,
\begin{eqnarray}
  \left((\om+\g)z\dd -\sqrt{\eta}g\left(z+\dd\right)\right)\bp_1 + \bD\bp_2&=\bE\bp_1,
  \label{eff1}\\ 
  \left((\om+\g)z\dd +\sqrt{\eta}g\left(z+\dd\right)\right)\bp_2 + \bD\bp_1&=\bE\bp_2,
  \label{eff2}
\end{eqnarray}
with
\beq
\bD=\frac{\eta+1}{2}\D+\frac{\eta-1}{2}E,\qquad
\bE=\frac{\eta+1}{2}E+\frac{\eta-1}{2}\D.
\label{renorm}
\eeq
Obviously, the system (\ref{eff1},\ref{eff2}) corresponds to the usual quantum 
Rabi model (\ref{hamr}) with $\s_x$ and $\s_z$ interchanged and renormalized parameters,
\beq
\om \ra \om+\g,\quad g \ra \sqe g,\quad E\ra \bE,\quad \D \ra \bD.
\eeq
For $\g=0$, Equations (\ref{eff1},\ref{eff2}) are equivalent to (\ref{orig1},\ref{orig2}).
Division by $\om+\g$ leads to the familiar form \cite{braak2011}
\begin{eqnarray}
  (z-\tg)\dd\bp_1 -(\tg z +\tE)\bp_1 &= -\tD\bp_2,
  \label{old1}\\
  (z+\tg)\dd\bp_2 +(\tg z -\tE)\bp_2 &= -\tD\bp_1,
  \label{old2}
\end{eqnarray}
with
\beq
\tg=\frac{g}{\sqrt{\om^2-\g^2}}, \quad
  \tE= \frac{1}{\om^2-\g^2}(\om E+\g\D), \quad
  \tD=  \frac{1}{\om^2-\g^2}(\om \D+\g E).
\eeq
We note that the coupling constant $g$ is renormalized by the factor $(\om^2-\g^2)^{-1/2}$ which diverges for 
$\g\ra \om_-$, although less so than the renormalization of energy $E$ and qubit splitting $\D$, 
which are mixed for $\g\neq 0$.
Using the manifest $\Zz_2$ invariance of (\ref{old1},\ref{old2}) and adopting the G-function formalism developed by one of us in \cite{braak2011}, we obtain the $G$-functions 
\beq
G_\pm(x)=\sum_{n=0}K_n(x)\left(1\mp\frac{\tD}{x-n}\right)\tg^n,
\label{g-function}
\eeq
for each parity, with
the spectral parameter $x=\tE+\tg^2=(\om E+\g\D+g^2)/(\om^2-\g^2)$. The $K_n$ are 
determined recursively, $K_n=(f_{n-1}K_{n-1}-K_{n-2})/n$, with
\beq
f_n(x)=2\tg+\frac{1}{2\tg}\left(n-x+\frac{\tD^2}{x-n}\right).
\label{f-function}
\eeq
The distance between the poles at $E^{(p)}_n$ and $E^{(p)}_{n-1}$ corresponding to integer $x$ 
is $\delta E=(\om^2-\g^2)/\om$, in accord with the results in \cite{eckle2017,eckle2023,xie2019}. 
This distance tends to zero for $\g\ra\om_-$, when the Stark-coupling $\g$ becomes critical, 
similar to the two-photon QRM \cite{braak2023} for $g=\om/2$. In the latter case the spectrum 
develops a continuous part at the critical coupling. 
This continuous part of the spectrum could be inferred from the 
$G$-conjecture \cite{braak2011}, which assumes that the distribution of zeroes of $G_\pm(E)$ is to some extend determined by its poles. It reads,\par
\vspace{2mm}
\noindent
{\bf $\bm{G}$-conjecture:}\ Let $\{W_j| j\in\Zpl\}$ be the set of (simple) poles of $G(E)$. The zeroes of $G(E)$, forming the regular spectrum, $\{E_k|k\in\Zpl\}$ are either located below the first pole, $E_k<W_0$, or between two poles, $W_j < E_k < W_{j+1}$. In the latter case, there are at most two such zeros in the open interval $]W_j,W_{j+1}[$, i.e. $W_j < E_{k_l} < E_{k_l+1}<W_{j+1}$. If this is the case, the intervals $]W_{j-1},W_j[$ and $]W_{j+1},W_{j+2}[$ may contain either none or only one $E_k$. If the interval $]W_j,W_{j+1}[$ contains none of the $E_k$, the adjacent intervals $]W_{j-1},W_j[$ and $]W_{j+1},W_{j+2}[$ contain at least one $E_k$.\par
\vspace{2mm}
\noindent
This conjecture has been seen to be valid by extensive numerical checks in case of the QRM \cite{linh2024}. Very recently, the conjecture has been proven to be valid for almost all intervals $]W_j,W_{j+1}[$ \cite{rudnick2023}. 
The conjecture entails that at least one zero of $G_\pm(E)$ is located between $W_j$ and $W_{j+2}$. If the distance between the poles approaches zero as one of the parameters approaches a critical value, the discrete eigenvalues become more and more dense on the real axis above $W_0$, eventually forming a \emph{continuum} if the critical point is reached.

This happens also in the RS model as $\delta E$ goes to zero for $\g \ra\om_-$. Thus, 
one would deduce a continuum of non-normalizable states at $\om=\g$, extending from a threshold 
given by the position of the first pole at
$E_{c}=-(\D+g^2/\om)$ up to positive infinity. However, $E_{c}$ provides only 
an upper bound of the actual threshold energy, because the qualitative behavior of the zeroes of $G_\pm(E)$ 
below the first pole cannot be deduced in a simple manner.

From Equation (\ref{g-function}) we see that $G_+(x)=G_-(x)$ for $\tD=0$ or $E=-\om\D/\g$. If this energy happens to 
be an eigenvalue, it must be doubly degenerate. Indeed, the case $\tD=0$ corresponds to a degenerate 
qubit in the renormalized model, therefore $\tE+\tg^2=n$ with $n=0,1,2,\ldots$ 
which leads to the array of special crossing points at the same energy $E=-\om\D/\g$ independent from $g$ in 
the spectral graph. Compare Fig.2 in \cite{xie2019} and Equation (\ref{gcn}) above \cite{note3}. 
In contrast to the ordinary QRM, there is a crossing on the first baseline $x=n=0$, corresponding to 
$E=-(\g\D+g^2)/\om$, which translates to the condition on the parameters $g^2=\D(\om^2-\g^2)/\g$, which 
can be satisfied for $\g>0$ (if $\D>0$ as usual) \cite{xie2019}. If $\D$ is sufficiently small, 
this crossing corresponds to a degeneracy of the ground state with the first excited state 
and thus to a quantum phase transition of first order, as pointed out in \cite{xie2019}. 
We observe that the location of this phase transition will occur for small coupling $g$ 
(within the reach of experimental platforms) if $\g$ is close enough to $\om$.


\section{The limit $\bm{\g\ra\om_-}$}\label{sec-limit}

We consider the limit $\g\ra\om_-$ and start with Equations (\ref{old1},\ref{old2}). For $\g<\om$, 
the two regular singular points of this system are located at $z=\pm\tg=g/\sqrt{\om^2-\g^2}$. If $\g$ approaches 
$\om$ from below, these singular points tend to infinity to coalesce at $z=\infty$ for $\g=\om$, 
forming there an irregular singular point of s-rank three \cite{slavyanov2000}. 
In terms of the ODE's (\ref{old1},\ref{old2}), this process is called \textit{confluence} of the 
two singularities and  
well known e.g. from the hypergeometric equation, where the confluence of two regular singular 
points at infinity produces the Bessel equation \cite{slavyanov2000}. 
The first step in deriving the confluent equation is elimination of $\bp_2$ in terms of $\bp_1$. 
We apply the operator on the left hand side of (\ref{old2}) to (\ref{old1}) and obtain
\beq
\left(z^2-\tg^2\right)\bp_1''+\left(z\left[1-2\tg^2-2\tE\right]+\tg\right)\bp_1'
+\left(\tE^2-\tD^2-\tg^2-\tg z(1+\tg z)\right)\bp_1=0.
\label{confl1}
\eeq
The parameters in (\ref{confl1}) diverge as a power of $\eps^{-1}$ with $\eps=\sqrt{\om^2-\g^2}$. 
For example, $\tg^2$ diverges as $\eps^{-2}$, as do $\tE$ and $\tD$. 
The maximally divergent terms in Equation (\ref{confl1}) seem to be $\tE^2$ and $\tD^2$, diverging like $\eps^{-4}$. 
The confluence process consists now in letting $\eps \ra 0$ while keeping the most divergent terms. 
In our case these are actually $\sim \eps^{-2}$ because
\beq
\tE^2-\tD^2=\frac{E^2-\D^2}{\om^2-\g^2}.
\eeq
In the confluence process it is assumed that powers of $z$ multiplying $\bp_1(z)$ 
and its derivatives are bounded with respect to the diverging parameters. The result reads
\beq
-g^2\bp_1''-2(g^2+\om E+\om\D)z\bp_1' +\left(E^2-\D^2-g^2-g^2z^2\right)\bp_1=0,
\eeq
or
\beq
\left[\frac{1}{2}\frac{\rd^2}{\rd z^2}
  +\left(1+\frac{\om(E+\D)}{g^2}\right)\left[z\dd+\frac{1}{2}\right]
  +\frac{z^2}{2}\right]\bp_1=\Lambda\bp_1,
\label{confl2}
\eeq  
with
\beq
\Lambda=\frac{E^2-\D^2}{2g^2}+\frac{\om}{2g^2}(E+\D).
\label{Lambda}
\eeq
The form of Equation (\ref{confl2}) shows it is the eigenvalue problem of an operator which 
can be written as linear combination of the elements of $\mathfrak{sl}_2(\Rr)$, acting in $\B$ as \cite{vourdas2006},
\beq
K_-=\frac{1}{2}\frac{\rd^2}{\rd z^2}, \quad
K_0=z\dd+\frac{1}{2}, \quad
K_+=\frac{1}{2}z^2.
\eeq
We have
\beq
\left(K_- +\a K_0 +K_+\right)\bp_1=\Lambda\bp_1, \quad \a=1+\frac{\om}{g^2}(E+\D).
\label{sl2}
\eeq
As the $s$-rank of (\ref{confl2}) is three, the formal solutions have the following asymptotic form
\beq
\bp_1(z)=\exp\left(\frac{\b_1}{2}z^2+\b_2z\right)z^\rho\sum_{n=0}c_nz^{-n}, \qquad z\ra\infty.
\label{asym}
\eeq
Plugging the expansion (\ref{asym}) into (\ref{confl2}), we find for $\b_1$,
\beq
\b_1^{\pm}=-\a\pm\sqrt{\a^2-1}.
\label{beta1-l}
\eeq
This means that $\b_1$ is real with $|\b_1^+|<1$  and $|\b_1^-|>1$ for $\a>1$ and
vice versa for $\a<-1$. In this case, $|\a|>1$, one exponent of second kind has absolute value less than 1, 
which means that the corresponding solution is normalizable in $\B$. This situation corresponds thus to a discrete 
spectrum,
\beq
\begin{array}{ccc}
  \a>1 & \Leftrightarrow & E >-\D,\\
  \a<-1 & \Leftrightarrow & E< -\D-\frac{2g^2}{\om}.
  \end{array}
\eeq
For $|\a|<1$, the exponent $\b_1$ lies on the unit circle,
\beq
\b_1^\pm=-\a \pm i\sqrt{1-\a^2},
\eeq
and it depends on $\b_2$ and $\rho$, whether $\bp_1(z)$ is normalizable or not \cite{braak2023}. 
In our case, we find for $\b_2$ the equation,
\beq
\b_2\sqrt{\a^2-1}=0,
\eeq
which entails $\b_2=0$ for $\a\neq \pm 1$. Exactly at $\a=\pm1$, the exponents $\b_1$ become degenerate, 
either $\b^\pm_1=-1$ or $\b^\pm_1=1$, which hints at the special role of 
this case. $E=-\D$ ($\b_1^{\pm}=-1$) is the limit of $E=-\om\D/\g$, the energy corresponding to the array of special solutions mentioned in section \ref{sec-qrm}. 
For $|\a|<1$, 
the normalizability of $\bp_1(z)$ is determined by $\rho$,
\beq
\rho=-\frac{1}{2}-\frac{i}{\sqrt{1-\a^2}}\Lambda,
\eeq
independent from the value of $\b_1$. The fundamental property of
asymptotic expansions like (\ref{asym}) is the appearance of the Stokes phenomenon: A certain combination of exponents $\{\b_1,\b_2,\rho\}$ is only valid in a certain sector of the complex plane, centered around the (Stokes) ray $z(t)=te^{i\psi}$  where $\bp_1(z)$ is either dominant or recessive \cite{slavyanov2000}. In general, the analytic continuation of a solution having an asymptotic expansion with exponents $\{\b_1,\b_2,\rho\}$ in a given sector will have different exponents in the other sectors.
In our case, the Stokes rays have the form $z=\pm|z|e^{-i\th/2}$ for $\b_1=e^{i\th}$ if the solution has dominant 
asymptotics, $\bp_1(z)\sim\exp(|z|^2/2)$, in the sector $-\th/2-\pi/4<\textrm{arg}(z)<-\th/2+\pi/4$. Dominant asymptotics means in the present case, $|\b_1|=1$, 
that the Bargmann norm of $\bp_1$ is finite if and only if the integral
\beq
I_0(R) = \int_R^\infty\rd x\int_{-x}^x\rd y\
(x^2+y^2)^{\Re(\rho)}e^{-2\Im(\rho)\arctan(y/x)} e^{-2y^2}
\label{integral}
\eeq
is finite \cite{braak2023}. 
Here we have used the unitary transformation $z\ra e^{i\th/2}z$ to set $\b_1=1$, 
which means that the critical sectors, where this case may occur, are $S_0=-\pi/4<\textrm{arg}(z)<\pi/4$ 
and $S_3=3\pi/4<\textrm{arg}(z)<5\pi/4$ ($I_0(R)$ refers to $S_0$). The parameter $R$ must be 
large enough to ensure validity of the asymptotic expansion in the given sector to the supposed precision. 
We have $\Re(\rho)=-1/2$ which is, according to the analysis in \cite{braak2023}, the marginal case. Marginal means here that the solution $\bp_1(z)$ is normalizable if $\Re(\rho)<-1/2$ and not normalizable for $\Re(\rho)>-1/2$ \cite{braak2023}. 
To see this, let's look at the integration over $y$ in (\ref{integral}),
\begin{align}
&\int_{-x}^x\rd y\ (x^2+y^2)^{\Re(\rho)}e^{-2\Im(\rho)\arctan(y/x)}e^{-2y^2}   \nn\\
&= x^{2\Re(\rho)+1}\int_{-1}^1\rd \bar{y}(1+\bar{y}^2)^{\Re(\rho)}
e^{-2\Im(\rho)\arctan(\bar{y})}e^{-2x^2\bar{y}^2} \label{integral2}\\
&\approx x^{2\Re(\rho)+1}\frac{\sqrt{\pi/2}}{x} = \sqrt{\frac{\pi}{2}}x^{2\Re(\rho)}, \nn
\end{align}
because the saddle point in (\ref{integral2}) is very close to $\bar{y}=0$ for $x\ge R$ large enough. 
This leads to
\beq
I_0(R)\approx \sqrt{\frac{\pi}{2}}\int_R^\infty\rd x\ x^{2\Re(\rho)}.
\eeq
This integral converges for $\Re(\rho)< -1/2$ and diverges logarithmically for $\Re(\rho)=-1/2$. 
The formal solution $\bp_1(z)$ has no finite Bargmann norm for $|\a| < 1$, in accord with the 
result in \cite{maciejewski2015}. 
There are no isolated eigenvalues in the energy range $-\D-2g^2/\om < E < -\D$. 
Now we have two options:\par
\noindent
1)\ The range $-\D-2g^2/\om < E < -\D$ is a spectral gap and does not contain any eigenvalues.\\
2)\ There is a spectral continuum between $-\D-2g^2/\om$ and $-\D$.\\
To decide between these two possibilities, it is not enough to state merely the 
non-normalizability of the entire formal solutions of (\ref{confl2}) as in \cite{maciejewski2015}  
because these formal solutions are entire (and non-normalizable) also in the gaps between the discrete eigenvalues 
for $|\a|>1$. 

To show that there is indeed a continuum, we must demonstrate that the formal eigenfunctions $\bp_\Lambda(z)$
for eigenvalue $\Lambda$ furnish a spectral measure on the real line absolutely continuous with respect to 
Lebesgue measure \cite{reed-1}. For example, the formal eigenfunctions 
$\p_p(x)=(2\pi)^{-1/2}e^{ipx}$ of the momentum operator $-i\rd/\rd x$ satisfy the 
generalized orthonormality relation,
\beq
\langle\p_p|\p_{p'}\rangle=\delta(p-p')
\label{scalarp}
\eeq
and furnish thus a proper spectral measure on the real line $-\infty < p < \infty$, 
because any element of $L^2(\Rr)$ can be expanded in terms of the set $\{\p_p(x)\}$ as a Fourier integral. 
In $\B$, the functions $\p_p$ read \cite{braak2023},
\beq
\p_p(z)=\pi^{-1/4}e^{-\frac{p^2}{2}}\exp\left(\frac{z^2}{2}+i\sqrt{2}pz\right).
\eeq
These functions have the asympotics (\ref{asym}) with $\b_1=1$, $\b_2=i\sqrt{2}p$ and $\rho=0$ everywhere in the complex plane, there is no Stokes phenomenon. The Stokes rays with dominant asymptotics (the critical lines) are the positive and negative real axis, the corresponding critical sectors being thus $S_0$ and $S_3$ as above.
Our case is somewhat similar, apart from the fact that we know only the 
asymptotics and not the solution itself in closed form. Moreover, the critical lines of $\bp_\La(E)$ depend
on $\La(E)$ via $\a(E)$ and we don't know a priori whether the Stokes phenomenon occurs or not. If the asymptotics in a critical sector belonging to $\b_1(\La(E))$ do \emph{not} contain the exponent $\b_1(\La(E))$, the solution will be normalizable. However, this normalizability will occur only for discrete set of $\La$'s, if at all. In the generic case, $\b_1(\La(E))$ will appear in the asymptotics of the corresponding critical sector, and this is the case treated below. 
If we write
\beq
\bp_{\La}(z) = \psi_{\La}(z) +C(\La)\exp\left(\frac{e^{i\th(\La)}z^2}{2}\right)z^{\rho(\La)},
\label{resol}
\eeq
where $\psi_{\La}(z)$ is a normalizable correction to the asymptotics in the sector 
containing the critical line $z=|z|e^{-i\th/2}$,
it is clear that $\langle \bp_{\La_1}|\bp_{\La_2}\rangle$ vanishes for $\La_1\neq\La_2$ due 
to the hermiticity of the operator on the left hand side of (\ref{confl2}). To investigate the 
orthonormality relation (\ref{scalarp}), we can therefore restrict to $\La_1=\La_2+\eps$ with small $\eps$. 
Without changing the result, we may thus assume $\th(\La_1)=\th(\La_2)$ and transform it to zero as above.
The relevant contributions to  $\langle \bp_{\La_1}|\bp_{\La_2}\rangle$ come now from the integral over 
sectors $S_0$ and $S_3$. We have in $S_0$ with $z=x+iy$,
  \begin{align}
\langle \bp_{\La_1}|\bp_{\La_2}\rangle &\propto 
\int_{S_0}\rd x\rd y \ e^{-x^2-y^2}e^{\frac{1}{2}\Big((x+iy)^2+(x-iy)^2\Big)}(x-iy)^{\rho^\ast(\La_1)}(x+iy)^{\rho(\La_2)} \\
&= \int_0^\infty\rd x\  x^{\rho^\ast(\La_1)+\rho(\La_2)+1}\int_{-1}^1\rd \bar{y}
\ (1-i\bar{y})^{\rho^\ast(\La_1)}(1+i\bar{y})^{\rho(\La_2)}e^{-2x^2\bar{y}^2}. \nn
  \end{align}
  The integral over $\bar{y}$ depends non-trivially on $x$ for small values of $x$, 
  especially the divergence at $x=0$ of the estimate in (\ref{integral2}) will be lifted. 
  However, the corresponding deviations could as well be absorbed in the normalizable 
  terms $\psi_{\La_1}(z), \psi_{\La_2}(z)$ which do not influence the orthonormality relation. Therefore, we may write
  \begin{align}
  \langle \bp_{\La_1}|\bp_{\La_2}\rangle &= C\int_0^\infty\frac{\rd x}{x}
  \ x^{i\big(\Im(\rho(\La_2))-\Im(\rho(\La_1))\big)} \nn\\
  &= C\int_{-\infty}^{\infty}\rd t\ \exp\left(it\left[\frac{\La_1}{\sqrt{1-\a_1^2}}-\frac{\La_2}{\sqrt{1-\a_2^2}}\right]\right)\\
  &\propto\delta(\La_1'-\La_2'),\nn
  \end{align}
  with $\La'=\La(1-\a^2)^{-1/2}$. 
  It follows that the generalized eigenfunctions $\bp_\La$ furnish indeed a proper spectral measure of $H_{RS}$ 
  at the critical point $\om=\g$ in the energy interval $[-\D-2g^2/\om,-\D]$, so that there is no gap but a 
  continuum of (generalized) states present.

  We return to the case $|\a|>1$, where a discrete spectrum exists. At first sight, its determination is 
  complicated by the fact that there is only a single irregular point at infinity, similar to the two-photon QRM. However, in contrast to that case, the equation for $\bp_1$ containing the $\mathfrak{sl}_2(\Rr)$ elements 
  is not coupled to another equation for $\bp_2$. The operator in (\ref{sl2}) is itself an element of  $\mathfrak{sl}_2(\Rr)$ and corresponds in the defining two-dimensional representation to the matrix
  \beq
  A=\left(\!\!
  \begin{array}{cc}
    \a & 1\\
    -1&-\a
  \end{array}\!\!\right).
  \label{slmatrix}
  \eeq
  First, we study $\a>1$. The matrix $A$ is diagonalized by the transformation
  \beq
  U=\frac{1}{\sqrt{1-a^2}}
  \left(\!\!
  \begin{array}{cc}
    -1 & a\\
    a&-1
  \end{array}\!\!\right),
  \qquad
  U^{-1}AU=
  \left(\!\!
  \begin{array}{cc}
    \sh\th & 0\\
    0&-\sh\th
  \end{array}\!\!\right),
  \label{transf}
  \eeq
  with $a=\ch\th-\sh\th$ and $\ch\th=\a$. It follows with $a<1$ that $U$ is an element of $SU(1,1)$, 
  a real subgroup of $SL_2(\Cc)$ which acts on $\B$ as isometries \cite{duan2022,braak2023}. 
  This means that there is a unitary transformation in $\B$ which maps the left hand side of (\ref{sl2}) 
  onto $\sh\th K_0$ which is the Hamiltonian of the harmonic oscillator with frequency $\sh\th$ 
  (including the vacuum term). It follows at once that the eigenvalue equation (\ref{sl2})
  is equivalent to
  \beq
  \sqrt{\a(E)^2-1}\left(n+\frac{1}{2}\right) =\La(E), \quad n=0,1,2,\ldots
  \label{harm1}
  \eeq
  in accord with \cite{maciejewski2015,xie2019}. 
  The energy $E$ is given implicitly in (\ref{harm1}) and leads to a lower bound for the discrete spectrum with $\a>1$, namely $E>$max$(-\D,\D-\om)$. 
  In the case $\a<-1$, the same reasoning must be slightly 
  modified because now $\ch\th=-\a$ and the
  transformation $U$ which yields the diagonal form in (\ref{transf}) has now $a=\sh\th +\ch\th>1$ and is not 
  an element of $SU(1,1)$. If one exchanges the two eigenvectors, one obtains
  \beq
  U=\frac{1}{\sqrt{a^2-1}}
  \left(\!\!
  \begin{array}{cc}
 a & 1\\
    1&a
  \end{array}\!\!\right),
 \eeq
 which belongs to $SU(1,1)$ and is therefore a valid isometry. 
 The eigenvalue equation reads for $\a<-1$,
 \beq
 \sqrt{\a(E)^2-1}\left(n+\frac{1}{2}\right)=-\La(E)
 \label{harm2}
 \eeq
 which yields together with (\ref{Lambda}) the \textit{same} inequality on $E$ as (\ref{harm1}), namely
 \beq
 E>\D-\om.
 \label{cond1}
 \eeq
 This restricts the possible values for $E<-\D-2g^2/\om$ considerably, 
 as well as the parameter regime where the lower discrete spectrum can exist at all. 
 The condition reads
 \beq
 \om\left(\frac{\om}{2}-\D\right)>g^2,
 \label{cond2}
 \eeq
 again in accord with \cite{maciejewski2015,xie2019}. 
 Beyond the coupling regime which supports a lower discrete spectrum, the continuum 
 threshold at $E_{thr}=-(\D+2g^2/\om)$ is at the same time the ground state energy. 
 As $E_{thr}<E_c$, the lower limit of the continuum deduced from the pole structure of the $G$-function, 
 we see that the confluence process going from (\ref{confl1}) to (\ref{confl2}) is consistent with the 
 continuum predicted from  $G_\pm(x)$, at least regarding the energy region $E<-\D$. On the other hand, 
 the $G$-function also predicts that the continuum should extend over the whole real axis from $E_c$ upwards. 
 How can this be reconciled with the confluence process described in (\ref{confl1}) and (\ref{confl2})?

 We have already mentioned that the confluence process assumes the relative boundedness of the variable prefactors 
 like $z^2$ in (\ref{confl1}). We shall describe now an alternative confluence process, starting from (\ref{eff1}) 
 and (\ref{eff2}), which lifts this condition. We divide by $\eta$ and obtain, defining now $\eps=\sqrt{\om-\g}$,
 \begin{align}
  \left(\eps^2z\dd -\frac{g\eps}{\sqrt{\om+\g}}\left(z+\dd\right)  +\Gamma_-\right)\bp_1 + \Gamma_+\bp_2&=0,
  \label{confl31}\\
 \left(\eps^2z\dd +\frac{g\eps}{\sqrt{\om+\g}}\left(z+\dd\right)  +\Gamma_-\right)\bp_2 + \Gamma_+\bp_1&=0,
  \label{confl32}
\end{align}
 with
 \beq
 \Gamma_\pm=\frac{\D-E}{2(\om+\g)}\eps^2\pm \frac{\D+E}{2}.
 \label{gamma}
 \eeq
 The next step consists in a representation of $\ad, a$ (i.e. $z$  and $\rd/\rd z$) by position and 
 momentum operators as
 \beq
 a=\frac{1}{\sqrt{2}}\left(\zeta \tx +\frac{1}{\zeta}\frac{\rd}{\rd \tx}\right),
 \qquad
 \ad=\frac{1}{\sqrt{2}}\left(\zeta \tx -\frac{1}{\zeta}\frac{\rd}{\rd \tx}\right).
 \label{alta}
 \eeq
 This representation preserves the canonical commutation relations.
We set now $\zeta=\sqrt{2C}/\eps$, this means that $\zeta$ diverges as $\eps \ra 0$. In this limit, the transformation is merely ``pseudo-canonical", a phenomenon known from systems with infinitely many degrees of freedom \cite{segal1970}.
The constant $C$ has the dimension of energy. Performing the limit yields
 \begin{eqnarray}
  \left(C\tx^2-\sqrt{\frac{2C}{\om}}g\tx - \frac{E+\D}{2} \right)\bp_1 + \frac{E+\D}{2}\bp_2&=0,
  \label{confl41}\\
 \left(C\tx^2+\sqrt{\frac{2C}{\om}}g\tx - \frac{E+\D}{2} \right)\bp_2 + \frac{E+\D}{2}\bp_1&=0.
  \label{confl42}
\end{eqnarray}
 Apparently, both $\bp_1(\tx)$ and $\bp_2(\tx)$ are generalized eigenfunctions of the position operator $\tx$.
 Eliminating $\bp_2(\tx)$ gives
 \beq
 C\tx^2\left(C\tx^2-\frac{2g^2}{\om} -(\D+E)\right)\bp_1(\tx)=0.
 \label{confl5}
\eeq
This translates via $\tx^2\ge 0$ into a condition for $E$,
\beq
E\ge -\D-\frac{2g^2}{\om}=E_{thr}.
\label{Ethr}
\eeq
It follows that the whole real axis above $E_{thr}$ represents a spectral continuum of $H_{RS}$ 
at the critical point $\om=\g$. 
The generalized eigenstates with energy $E$ have the form $\psi_{\tx_0}(\tx)\propto\delta(\tx-\tx_0)$ with $\tx_0=\pm\sqrt{(E-E_{thr})/C}$. To compare these states with the normalizable eigenstates of (\ref{harm1}), we write $\psi_{\tx_0}$ as function of the original position operator $q=2^{-1/2}(a+\ad)=\zeta \tx$,
\beq
\psi_{\tx_0}(x)=\delta\left(q\pm \zeta\sqrt{(E-E_{thr})/C}\right).
\label{cont-2}
\eeq
For any finite energy $E > E_{thr}$, the generalized eigenstate will be centered around 
$q_\pm (E)=\pm \zeta\sqrt{(E-E_{thr})/C}$ which diverges like $1/\eps=(\om-\g)^{-1/2}$. The same is true of the two associated spectral measures on any interval $[E_1,E_2]$ with $E_1> E_{thr}$ which corresponds to elements of $L^2(\Rr)$ supported respectively on the intervals $\pm[q_+(E_1),q_+(E_2)]$. As the normalizable states of (\ref{harm1}) have wave functions exponentially decreasing for $q\ra\pm\infty$, we conclude that they will no longer hybridize with the states in (\ref{cont-2}) for $\g\ra\om_-$ and belong to a dynamically invariant subspace of $L^2(\Rr)$. Therefore, they can be considered as ``bound states in the continuum" at the critical point. However, for all $\g<\om$ the precursors of the BICs will hybridize with the precursors of the states in (\ref{cont-2}), leading to avoided crossings in the spectral graph which become more and more narrow as $\g$ approaches $\om$. True BICs exists only at the critical point itself, which is here defined as the limit $\g\ra\om_-$, not by setting $\g=\om$ in (\ref{orig1}) and (\ref{orig2}). The latter case was studied in \cite{maciejewski2015, xie2019} and leads just to the two discrete spectra (\ref{harm1}), (\ref{harm2}) and the ``small continuum",  obtained by the first confluence process given in (\ref{confl2}). The canonical transformation (\ref{alta}) can be implemented unitarily only if $\zeta<\infty$, so it fails exactly at $\g=\om$ which is a singular point of the transformation. This phenomenon bears close resemblance to the failure to implement the scale transformation (\ref{alta}) unitarily with $\zeta<\infty$ but for infinitely many bosonic variables \cite{segal1970}. The Rabi-Stark model at the critical point provides thus an analogue of the infrared divergences occurring in interacting quantum field theories -- just like the singular limit $\Delta/\om\ra\infty$ leads to a quantum phase transition of the QRM \cite{bakemeier2012,ashhab2013,hwang2015}, although the model contains only two degrees of freedom. The crucial difference being that in the latter case the model parameters have to take unphysical (infinite) values, whereas the Rabi-Stark model can be simulated within cavity QED right at the critical point $\g=\om$ \cite{grimsmo2013}. Our results entail that this critical point has to be regularized by setting $\g=\om-\eps^2$, before letting $\eps \ra 0$ to obtain the full spectrum at $\g=\om$.

\section{Numerical results for large-dimensional state space}\label{sec-num}
In this section we shall present numerical evidence for the appearance of the ``big" spectral continuum for $\g\ra\om_-$ which is obtained by the second confluence process. This continuum extends from the energy threshold $E_{thr}$ up to positive infinity. The precursors of the BICs with energy $E_{BIC}> $max$(-\D,\D-\om)$, obtained from the first confluence process, can be identified also for $\g<\om$ as lines ascending  in the spectral graph as function of $g$, forming narrow avoided crossings with the descending lines which eventually turn into the continuum at $\g=\om$. We call the precursors of the BICs for $\g<\om$ ``preBICs" and the discrete level lines associated with the emerging continuum ``preContinuum". In all examples below, we set $\om=1$ and consider positive parity only.

\begin{figure} [hbt!]
\begin{center}
\includegraphics[width=0.8\textwidth]{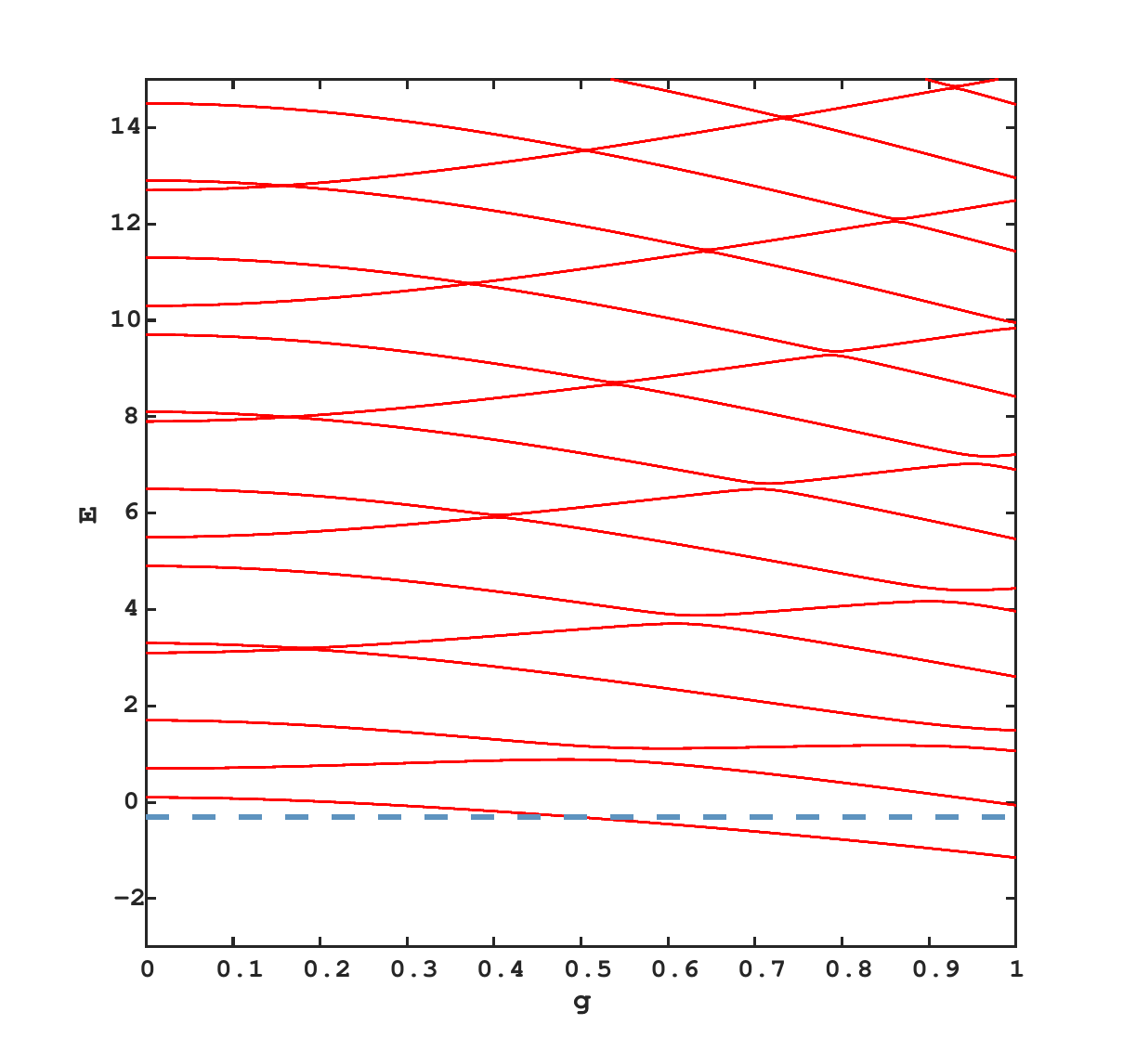}
\end{center}
\caption{
 $E(g)$ for $\Delta=0.7$, $\gamma =0.2$ (positive parity and $\omega=1$). The horizontal line at $E=\D-1$ gives the lower bound for the BIC energy as obtained from Equation (\ref{harm1}).} 
\label{fig-d.07-g.02}
\end{figure}

\begin{figure} [hbt!]
\begin{center}
\includegraphics[width=0.8\textwidth]{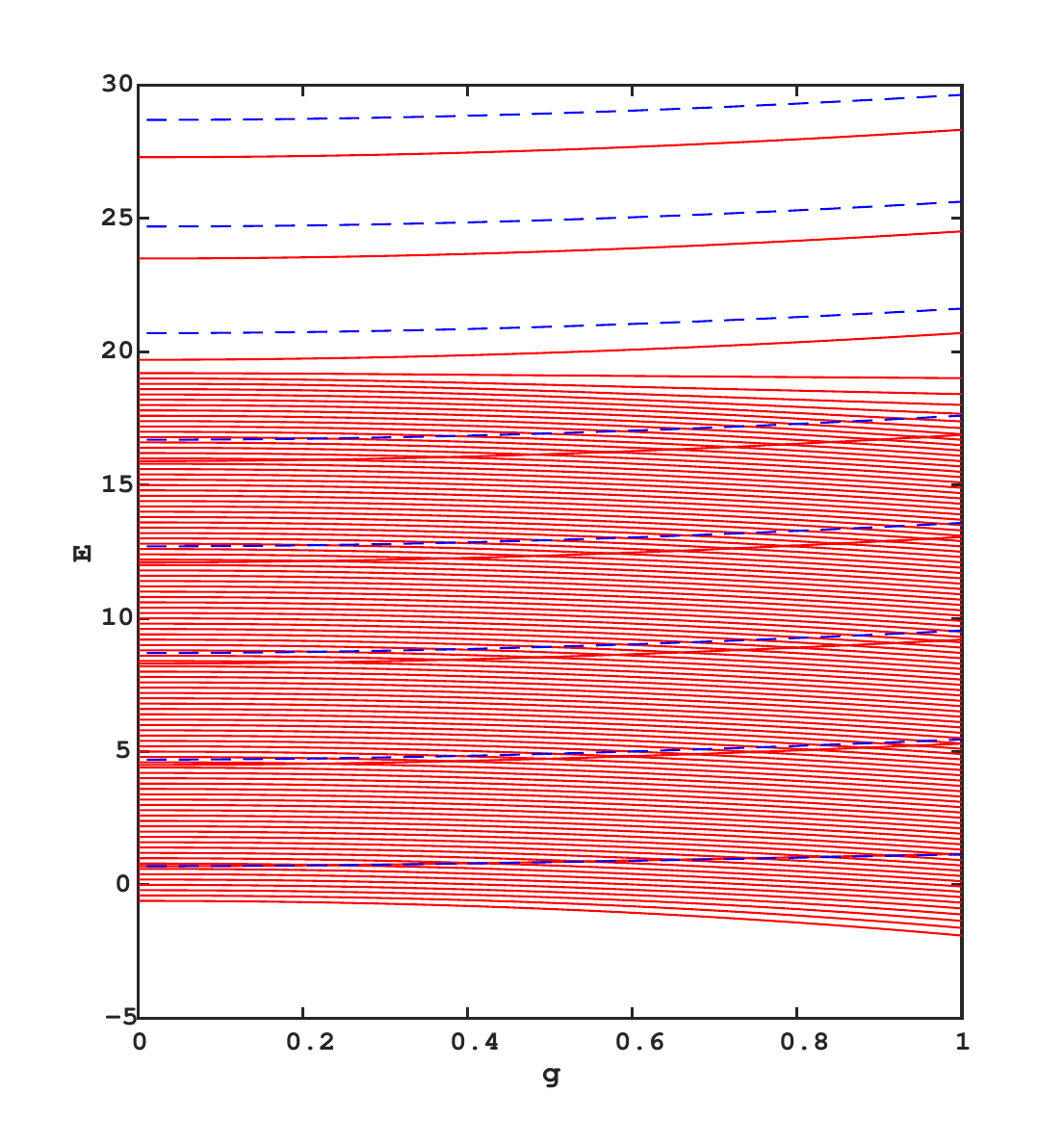}
\end{center}
\caption{
 $E(g)$ for $\Delta=0.7$, $\gamma =0.9$. The calculation uses 200 Fock states. The dashed blue lines indicate the even BIC spectrum for $\g\ra 1_-$. The BIC levels for $n=0, 2,\ldots 14$ are approximated by the corresponding preBIC levels, which show avoided crossings with the preContinuum for $n=0,2,\ldots 8$. The preBIC levels $n=10,12,14$ are continuous because they cannot hybridize with the preContinuum due to the low truncation number.} 
\label{fig-d.07-g.09}
\end{figure}

We employ the method of Exact Diagonalization (ED) to calculate the energy spectrum of the Rabi-Stark model, a proven technique for understanding the quantum properties of light-matter interaction models. ED is a powerful numerical method that involves solving the Hamiltonian matrix of the system directly to obtain its eigenvalues and eigenvectors. The robustness of ED, combined with its lack of reliance on any specific approximations, makes it an indispensable tool for the comprehensive exploration of the model. The high-performance computing capabilities of Vera C3SE provided the necessary computational power to carry out the extensive numerical analysis of our study, especially in regimes close to the critical point $\gamma \ra 1_-$.
Figure~\ref{fig-d.07-g.02} shows a case far away from the critical point, with $\D=0.7$ and $\g=0.2$.
We observe a spectral graph with two groups of ascending respectively descending lines. All crossings are (narrowly) avoided. As expected, the energy gap at the crossings grows with $g$ and becomes smaller for higher energy. The state space is truncated to 200 Fock states ($N_{trun}=200$) which suffices for the eigenvalues of the first 20 eigenstates to converge.
In Figure~\ref{fig-d.07-g.09}, with $\g=0.9$ the system is closer to the critical 
point ($\g\ra 1_-$) and shows already features reminiscent of the physics in this limit. 
The state space is still restricted to 200 Fock states, a truncation which is not sufficient to find all states below $E=30$. The three isolated states above the region of dense levels are preBICs (corresponding to the numbers $n=10,12,14$ in Equation (\ref{harm1})) and obtained in the numerical calculation because the true eigenstates with a preBIC energy have low photon content, \textit{i.e.}~they are located in a state space with low dimension, approximating the full Fock space $l^2(\Rr)$. But with $N_{trun}=200$ one cannot reach the states in the preContinuum which hybridize with the preBICs in Figure~\ref{fig-d.07-g.09} with numbers $10$ to $14$, therefore those preBIC level lines are continuous and show no avoided crossings, in contrast to the preBICs $0$ to $8$ which exhibit avoided crossings with the low-lying states of the preContinuum which is cut off at $E\sim 20$ due to $N_{trun}=200$. 
Remarkably, the preBIC $|\psi_0\rangle$ for $n=0$ is almost perfectly approximated by the BIC for $n=0$. This can be understood from the fact that this state has  photon content $\langle\psi_0|a^\dagger a|\psi_0\rangle \sim 0$, as well as the corresponding BIC. Because the Hamiltonian obtained from the first confluence process in Equation (\ref{confl2}) is equivalent to set $\g=\om$ in Equations (\ref{orig1}, \ref{orig2}) which removes the harmonic oscillator term in the lower spin sector, the difference between the BIC Hamiltonian in (\ref{confl2}) and the true one for $\g=0.9\om$ is very small if the photon content of the considered eigenstates is almost zero. For the same reason, the eigenvalues of BICs and preBICs deviate more strongly for higher photon content, as seen in Figure~\ref{fig-d.07-g.09}.

\begin{figure} [hbt!]
\begin{center}
\includegraphics[width=0.8\textwidth]{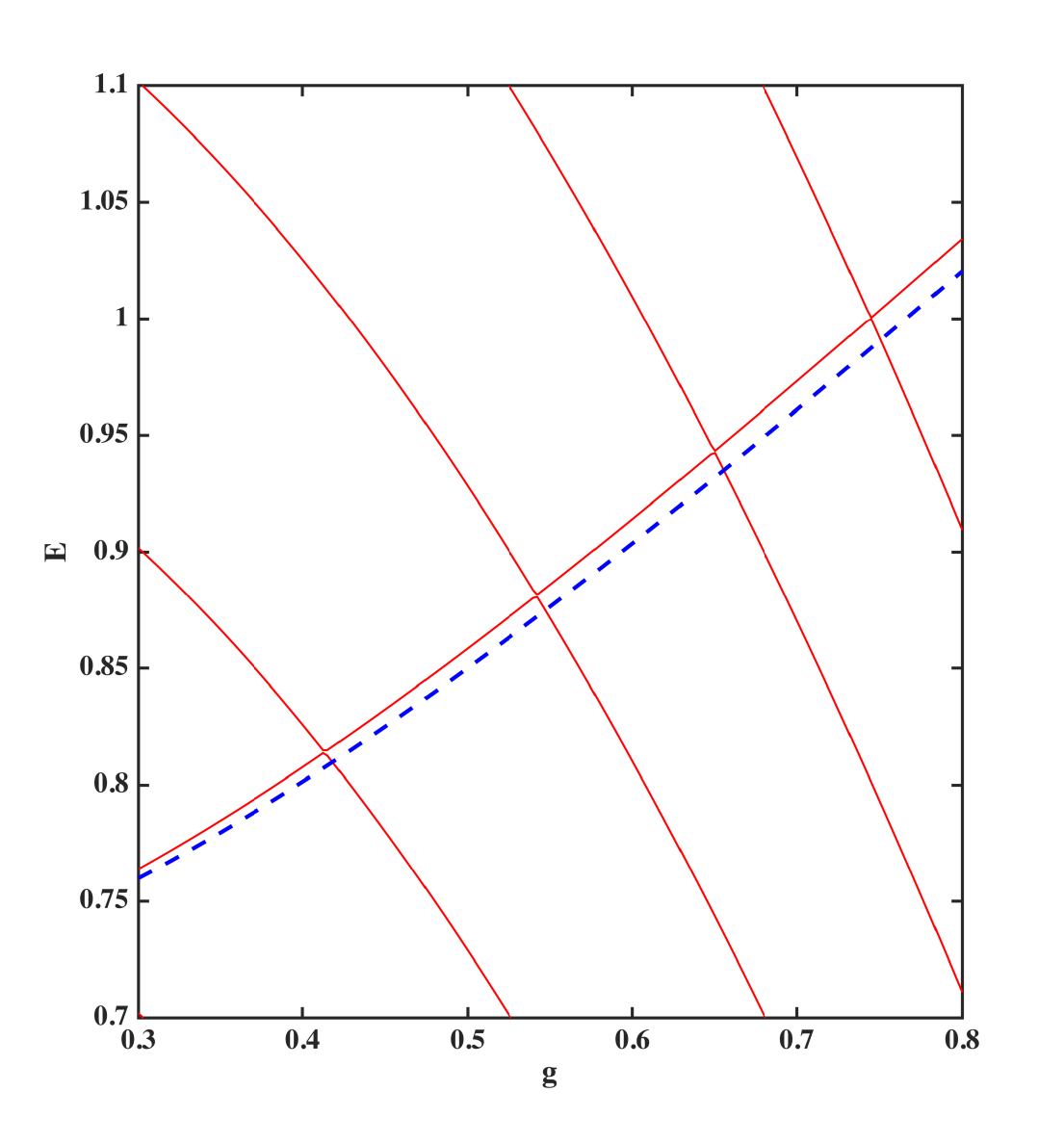}
\end{center}
\caption{Same parameters as Fig.\,\ref{fig-d.07-g.09}. The blue line indicates the BIC for $\g\ra 1_-$ and the ascending line the associated preBIC. It shows very small avoided crossings with the descending lines from the preContinuum.} 
\label{fig-3}
\end{figure}

The hybridization between preBIC and preContinuum is shown for $n=0$ in Figure~\ref{fig-3}. The avoided crossings are already very small.
It is evident from Figure \ref{fig-d.07-g.09} that enlarging the dimension of the state space beyond $N_{trun}=200$ will fill up the whole positive energy axis with level lines whose average spacing is given by the pole distance $\delta(E)=1-\g^2$ of the $G$-function $G_+(x(E))$, which is $\sim 0.2$ for $\g=0.9$. Compare the discussion after Equation (\ref{f-function}).

\begin{figure} [hbt!]
\begin{center}
\includegraphics[width=0.8\textwidth]{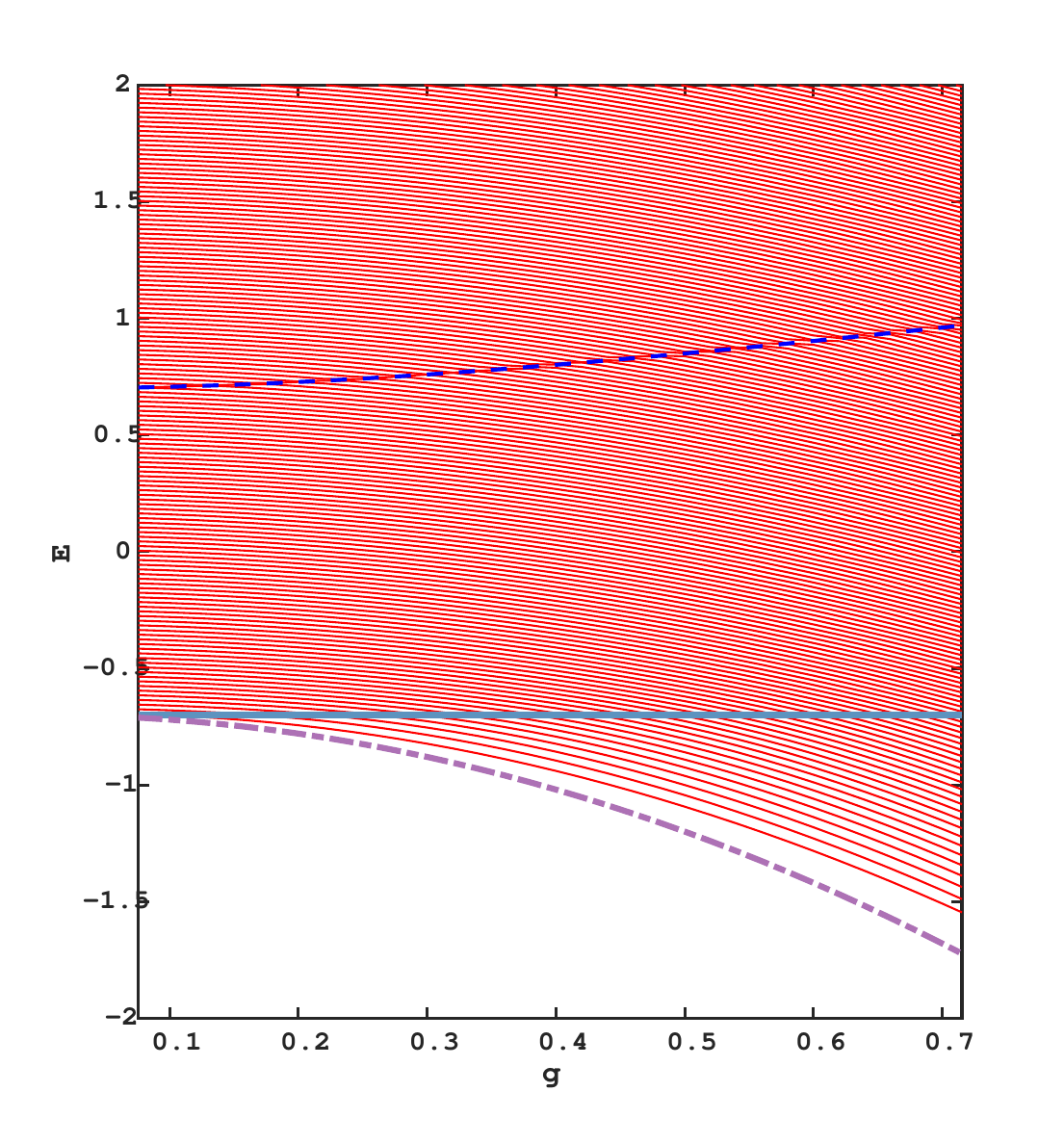}
\end{center}
\caption{E(g) for $\Delta=0.7$ and $\g=0.99$. $N_{trun}=500$. The dashed blue line indicates the first BIC and the dash-dotted magenta line $E_{thr}$. The density of levels in the preContinuum is smaller below $-\D$ (blue line) than above.} 
\label{fig-4}
\end{figure}

For $\D=0.7$, there exists no lower discrete spectrum at $\g=1$. So we do not expect to see the lower bound states (LBS) predicted by the first confluence process in this case. The ground state should eventually reach the threshold value $E=-\D-2g^2$ for $\g\ra 1_-$, without spectral gap to the excited states which form a continuum for both confluence processes. The case $\g=0.99$ is shown in Figure \ref{fig-4}. While the first preBIC is well approximated by the BIC with $n=0$ for $\g\ra1_-$, the ground state is separated by a gap from $E_{thr}$, which is bigger than the average level distance below $E=-\D$, the upper limit of the ``small" continuum predicted by the first confluence process. 

\begin{figure} [hbt!]
\begin{center}
  \includegraphics[width=0.8\textwidth]{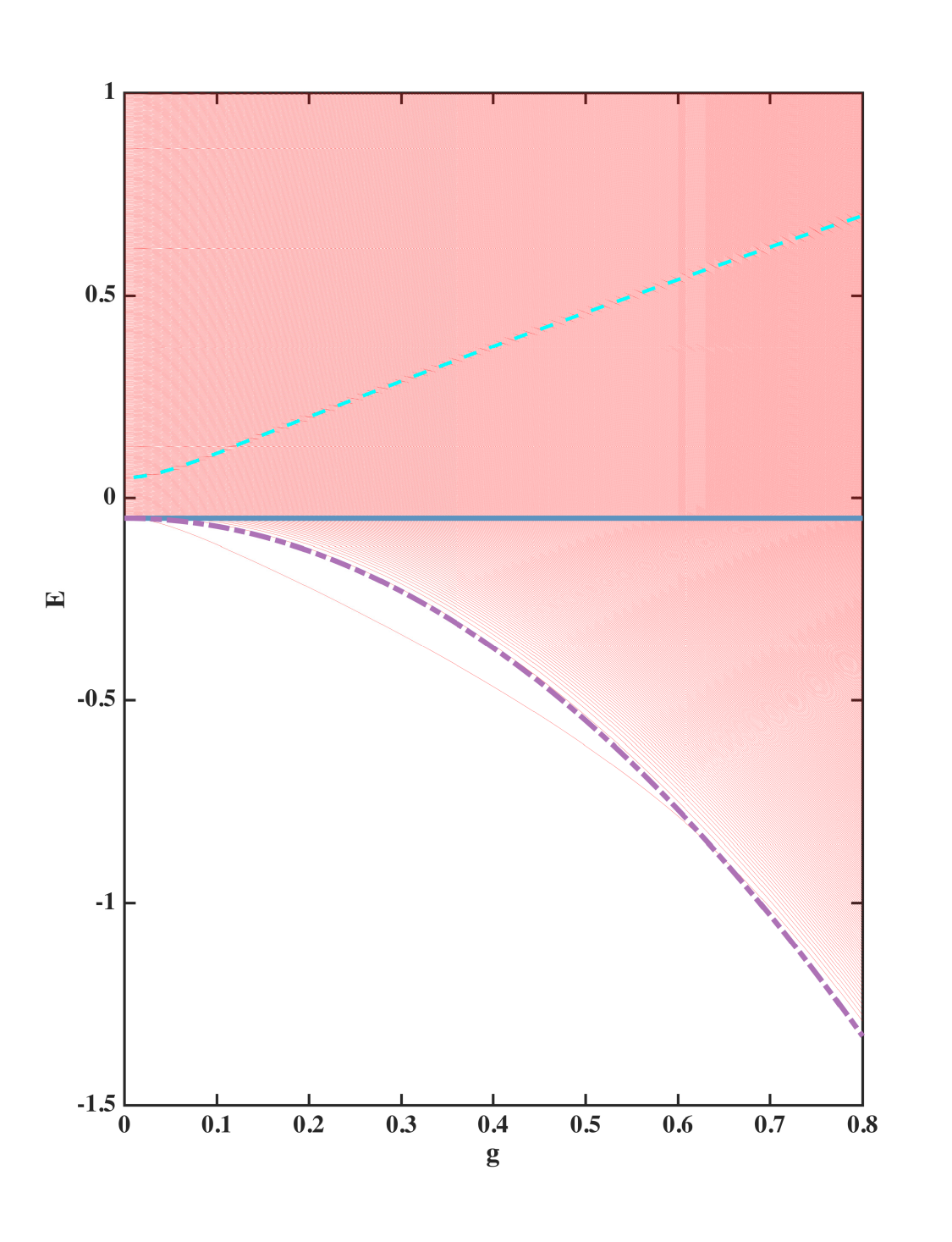}
\end{center}
\caption{E(g) for $\Delta=0.05$ and $\g=0.999$. $N_{trun}=10000$. The cyan dashed line indicates the first BIC (n=0), precisely on top of the associated preBIC. The dash-dotted magenta line denotes the threshold value $E_{thr}$ and the blue line the upper boundary of the ``small " continuum. At this value of $\g$, only one  precursor of the lower bound state (LBS) spectrum appears, the isolated ground state below the onset of the preContinuum (thin red line).} 
\label{fig-5}
\end{figure}

While the discrete spectra from the first confluence process give reasonable estimates for the preBICs in the case $\g=0.9$, we expect good approximations only for $\g$ very close to 1. In Figure (\ref{fig-5}), we depict the spectral graph for $\g=0.999$ and $\D=0.05$.
The state space dimension is here $N_{trun}=10000$. The area $-\D-2g^2<E<-\D$ marks the boundaries of the ``small" continuum according to the first confluence process. 
We see that the density of states shows no apparent change at $E=-\D$. 
This entails that the second confluence process predicting the ``big" continuum  for $E>E_{thr}$ in the limit $\g\ra 1_-$ is the correct description also for the preContinuum at $\g=0.999$ while the ``small" continuum seems to have no influence on the density of states for $\g\ra 1_-$. This question will be the subject of future study. Our results show furthermore that the set of lower bound states below threshold predicted by the first confluence process is only partially present in form of precursors even for $\g$ rather close to $\om$. Only one such precursor (the ground state) is visible in Figure~\ref{fig-5}.

\section{Analytical results in the slow-mode approximation}\label{sec-slowmode}

A more intuitive physical picture of how the BICs and the continuum arise may be obtained by examining an approximation that leads to a pair of effective potential bands for the coupled quantum system. This approach is adapted from a technique employed for the QRM in \cite{Graham1984} and analysed in more detail in \cite{EKIthesis}. Equation (\ref{hamrs}) may be written in terms of the position and momentum operators $\hat{q} = (2m \omega)^{-1/2}(\ad + a)$ and $\hat{p} = i(m \omega/2)^{1/2}(\ad - a)$ as
\beq
    H_{RS}= \frac{m \omega^2}{2}\left( 1 + \frac{\gamma}{\omega} \sigma_x \right) \hat{q}^2 + \frac{1}{2m} \left( 1 + \frac{\gamma}{\omega} \sigma_x \right) \hat{p}^2 - \frac{1}{2} \omega + \left(\Delta - \frac{\gamma}{2} \right) \sigma_x + g \sqrt{2m \omega} \hat{q} \sigma_z .
\eeq
Working in the position representation for the field, the state of the coupled system may be represented as
\beq
\ket{\psi} = \psi_+(q) \ket{+x} + \psi_-(q) \ket{-x} ,
\label{coupled_state}
\eeq
where $\sigma_x \ket{\pm x} = \pm \ket{\pm x}$. With $\hat{q} = q$ and $\hat{p} = -i d/dq$ in this basis, the Schr{\"o}dinger equations for the $\ket{\pm x}$
components become
\beq
\frac{m \omega^2}{2}\left( 1 \pm \frac{\gamma}{\omega} \right) q^2 \psi_{\pm}(q) - \frac{1}{2m} \left( 1 \pm \frac{\gamma}{\omega} \right)\frac{d^2}{dq^2}\psi_{\pm}(q) \pm \Dp \psi_{\pm}(q) + g \sqrt{2m \omega} q\, \psi_{\mp}(q) 
= \Ep \psi_{\pm}(q) ,
\label{scheq_pmx}
\eeq
where $\Dp \equiv \Delta - \gamma/2$ and $\Ep = E + \omega/2$. Neglecting the momentum term (cf. the second confluence prescription after Equation (\ref{alta})), a pair of coupled algebraic equations for the field components $\psi_{\pm}(q)$ is obtained:
\beq
\left[\frac{m \omega^2}{2} \left(1 + \frac{\gamma}{\omega} \right) q^2 \pm \Dp - \Ep \right] \psi_{\pm}(q) = -g \sqrt{2m \omega} q\, \psi_{\mp}(q) .
\label{comprel}
\eeq
Eliminating $\psi_{\pm}(q)$ and solving for $E$ leads to a pair of effective potential energy functions for the field degree of freedom,
\begin{align}
E_{a} &= \frac{m \omega^2}{2} q^2 - \frac{\omega}{2} + \left[ \left(\Delta - \frac{\gamma}{2} + \frac{m\omega \gamma}{2} q^2 \right)^2 + 2 m \omega g^2 q^2 \right]^{1/2} , \label{up_band}\\
E_{b} &= \frac{m \omega^2}{2} q^2 - \frac{\omega}{2} - \left[ \left(\Delta - \frac{\gamma}{2}  + \frac{m\omega \gamma}{2} q^2 \right)^2 + 2 m \omega g^2 q^2 \right]^{1/2} . \label{low_band}
\end{align}
Substituting $E_{a,b}$ for $\Ep$ in Equation (\ref{comprel}) gives the following relationships between the field components in the two energy bands:
\begin{align}
    \psi_+^a(q) &= \frac{g \sqrt{2m \omega} q}{\left[E_a - \Dp - \frac{m\omega^2}{2} \left(1 + \frac{\gamma}{\omega} \right) q^2 \right]} \psi_-^a(q) , \\
    \psi_-^b(q) &= \frac{g \sqrt{2m \omega} q}{\left[E_b + \Dp - \frac{m\omega^2}{2} \left(1 - \frac{\gamma}{\omega} \right) q^2 \right]} \psi_+^b(q) .
\end{align}
Inserting these back into the Schr{\"o}dinger equations (\ref{scheq_pmx}) and dividing through by $(1 \pm \gamma/\omega)$, as appropriate, we obtain 
\begin{align}
    \frac{E_a(q)}{\left(1+\frac{\gamma}{\omega}\right)} \psi_+^a(q) - \frac{1}{2m} \frac{d^2}{dq^2} \psi_+^a(q) &= \frac{\Ep}{\left(1+\frac{\gamma}{\omega}\right)} \psi_+^a(q), \label{effSch_up}\\
    \frac{E_b(q)}{\left(1-\frac{\gamma}{\omega}\right)} \psi_-^b(q) - \frac{1}{2m} \frac{d^2}{dq^2} \psi_-^b(q) &= \frac{\Ep}{\left(1-\frac{\gamma}{\omega}\right)} \psi_-^b(q).  \label{effSch_low}   
\end{align}
These equations represent effective Schr{\"o}dinger equations for the field components $\psi_+^a(q)$ in the potential $E_a(q)/(1 + \gamma/\omega)$ and $\psi_-^b(q)$ in the potential $E_b(q)/(1 - \gamma/\omega)$. Once the solutions to these equations are found, the components associated with the opposite spin states, $\psi_-^a(q)$ and $\psi_+^b(q)$, are determined by Equation (\ref{comprel}). Here, however, we are primarily concerned with the spectrum of the model.

\begin{figure} [hbt!]
\begin{center}
\includegraphics[width=1.\textwidth]{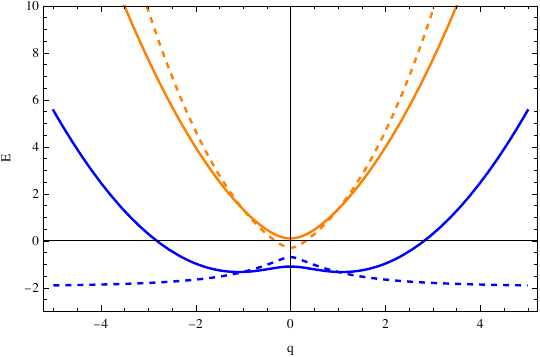}
\end{center}
\caption{Effective potential bands $E_a(q)$ (upper, orange) and $E_b(q)$ (lower, blue) for $\omega = 1, \Delta = 0.7$. Solid curves: $\gamma = 0.2$; dashed curves: $\gamma = 1$. } 
\label{fig:eff_pots}
\end{figure}

Simply studying the behavior of the effective potentials in Equations (\ref{up_band}) and (\ref{low_band}) provides considerable physical insight into the problem. Plots of these effective potentials are shown in Figure~\ref{fig:eff_pots}. The upper band, $E_a(q)$, has a single minimum at $q = 0$ for all parameter values (assuming $\Delta \ge 0$ and $0 \le \gamma \le \omega$) and goes to $+\infty$ as $q \to \pm \infty$. Hence it has a discrete energy spectrum and its eigenstates are normalizable bound states. To second order in $q$,
\beq
E_a(q) \approx \Dp - \frac{1}{2} \omega + \frac{m \omega^2}{2} \left(1 + \frac{\gamma}{\omega} + \frac{2g^2}{\omega \Dp} \right) q^2 .
\eeq
With this approximation, Equation (\ref{effSch_up}) reduces to the standard Schr{\"o}dinger equation for a one-dimensional quantum harmonic oscillator, with energy levels
\begin{equation}
    E_n^a = \Dp + (n+\tfrac{1}{2}) \omega_+ \left(1 + \frac{2 g^2}{\omega_+ \Dp} \right)^{1/2} - \frac{\omega}{2} 
    \label{harm_en_up} 
\end{equation}
where $n = 0,1,2,\dots$ and $\omega_+ = \omega(1 + \gamma/\omega)$. 
For fixed $n,\gamma,\D$, the energies increase with $g$, leading to ascending level lines in the spectral graph.
As $\gamma$ increases, the effective frequency of this harmonic potential also increases and the energy levels become more widely spaced.

\begin{figure} [hbt!]
\begin{center}
\includegraphics[width=1\textwidth]{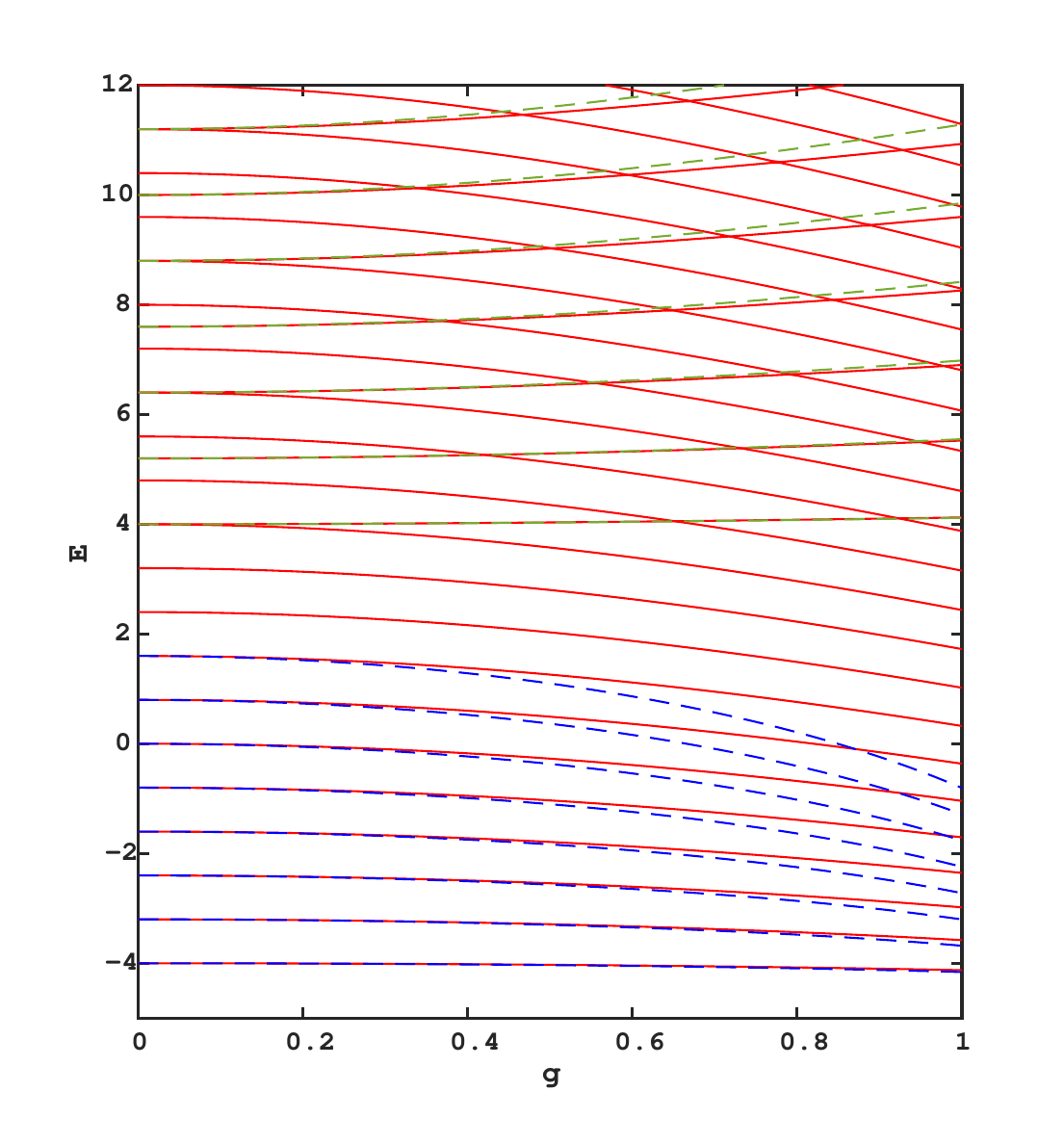}
\end{center}
\caption{In this spectral graph, both positive and negative parity of the numerical spectra are presented for $\D=4$ and $\g=0.2$. In the meanwhile, both $E_n^a$ (green) and $E_n^b$ (blue) according to Equations (\ref{harm_en_up},\ref{harm_en_low}) with $n=0,1,2, ... ,7$ are plotted.} 
\label{fig-7}
\end{figure}

The lower band, $E_b(q)$, displays more complicated behavior. Expanding in $q$,
\beq
E_b(q) = \Dp - \frac{1}{2} \omega + \frac{m \omega^2}{2} \left(1 - \frac{\gamma}{\omega} - \frac{2g^2}{\omega \Dp} \right) q^2 + \left(\frac{m \omega^2}{2}\right)^2 \left(\frac{2 g^4}{\omega^2 (\Dp)^3} + \frac{2 g^2 \gamma}{\omega^2 (\Dp)^2} \right) q^4 + \mathcal{O}(q^6).
\eeq
By keeping terms up to $\mathcal{O}(q^2)$, we again obtain a harmonic potential that can be solved exactly, with energies
\beq
    E_n^b = -\Dp + (n+\tfrac{1}{2}) \omega_- \left(1 - \frac{2 g^2}{\omega_- \Dp} \right)^{1/2} - \frac{\omega}{2} ,
    \label{harm_en_low}
\eeq
where $n = 0,1,2,\dots$ and $\omega_{-} = \omega(1 - \gamma/\omega)$. 
Formula (\ref{harm_en_low}) makes sense for $g$ below the value given in (\ref{g_c}) and leads there to descending level lines as function of $g$. Both behaviors are displayed in Figure~(\ref{fig-7}). This allows to interpret the ascending lines, which become preBICs for $\g\ra\om$, with the upper band (\ref{up_band}) and the descending lines, forming the preContinuum, with the lower band (\ref{low_band}).

For the lower band, increasing $\gamma$ decreases the effective frequency of the harmonic potential and the energy level spacing becomes smaller. However, at the point
\beq
\frac{2g^2}{\omega \Dp} = 1 - \frac{\gamma}{\omega} ,
\label{g_c}
\eeq
the harmonic term in the expansion vanishes. The lowest-order term then becomes
\beq
\left(\frac{m\omega^2}{2}\right)^2 \frac{1}{2(\Delta - \frac{\gamma}{2})} \left(1 - \frac{\gamma^2}{\omega^2} \right) q^4 .
\eeq
As $g$ increases above this  point, the potential changes from a single minimum at $q = 0$ to a double-well structure. For $0 \le \gamma < \omega$, the potential goes to infinity as $q \to \pm \infty$ and the spectrum remains discrete. However, it widens and flattens as $\gamma\ra\om$, so the spacing between eigenstates becomes smaller and smaller. At $\gamma = \omega$, the potential at $\pm \infty$ changes from $+\infty$ to a finite constant value:
\beq
\lim_{q \to \pm \infty} E_b(q) = -\Delta - \frac{2g^2}{\omega} = E_{thr}.
\eeq
This potential admits a continuum of non-normalizable scattering states, with a minimum energy equal to the threshold energy found above. As expected from the identification with descending level lines, the lower band produces a continuum for $\g\ra\om$ while the upper band still contains normalizable states which can be identified with the BICs from the first confluence process.

The effective potential approach is, of course, only approximate. As such, the regime of validity of the approximation must be considered. In this picture, the coefficients of the spin states depend upon the position coordinate $q$ of the field degree of freedom (Equation (\ref{coupled_state})). One way of looking at it is that the spin states can adiabatically follow a change in $q$, which suggests that the approximation should hold in the regime $\Delta \gg \omega$. Alternatively, neglecting the momentum term implies that the curvature of the wavefunction, $d^2 \psi/dq^2$, is small, which again suggests that $\omega$ should be small in some sense (this further suggests that the approximation may be better for lower-lying states in the potential). Another way of understanding the dependence on $\Delta$ is that the avoided crossings occur at resonances between states in opposite bands. As $\Delta$ increases, the number of energy levels below the bottom of the upper band increases, which means that resonances between the states are pushed out to higher and higher orders in the coupling $g$, which in turn suggests that the hybridization vanishes for $\g\ra\om$ and the bound states become true BICs.

The validity of the picture for large $\Delta/\omega$ is supported by examining the energy spectra of the QRM \cite{EKIthesis}. As $\Delta/\omega$ becomes larger, the avoided crossings in the spectrum become narrower and sharper. To the eye, two sets of apparently continuous curves emerge, one curving upwards and the other downwards as a function of $g$. Although the harmonic approximation does not fully capture this behavior, especially for the lower band where the range of $g$ values for which it holds is restricted by the change from a single well to a double well, the validity of the overall picture and of the harmonic approximation both improve as $\Delta$ increases. 

Intriguingly, a similar effect is obtained in the QRSM, even for relatively small values of $\Delta$, by increasing $\gamma$. For $g=0$, the Stark term shifts the frequency of the field from $\omega$ to $\omega(1 \pm \gamma/\omega)$, where the sign depends on the eigenstate of the spin. Therefore, increasing $\gamma$ decreases (increases) the effective frequency of the field associated with the lower (upper) spin state, as discussed in the introduction. Again, this has the effect of pushing the resonances to higher orders. From a perturbation theory standpoint, this causes the avoided crossings to have narrower energy splittings and extend over a smaller range of $g$ values.

The intuitive interpretation provided by the effective potential picture is consistent with the results in sections \ref{sec-limit} and \ref{sec-num}. The preBICs/BICs are identified with the bound states of the upper potential band. As $\gamma$ increases toward $\omega$, the lower band becomes wider and flatter, compressing the energy levels into a quasicontinuum. As long as $\gamma$ is less than $\omega$, the potential becomes infinite at $q = \pm \infty$, so the spectrum remains discrete and the eigenstates are normalizable. In the limit $\g\ra\om$, two things happen in parallel: The effective potential $E_b(q)$ of the lower band (\ref{low_band}) becomes flat for $q\ra\pm\infty$ and the effective mass $m/(1-\g/\om)$ becomes infinite (see Equation (\ref{effSch_low})), freezing the particle at an arbitrary given position. The latter reasoning produces precisely the ``small" continuum with energy values $E_{thr}<E<-\D$ (note that max$(E_b(q))=E_b(0)=-\D$ for $\g=\om$), while the asymptotically flat band corresponds to scattering states with arbitrary high energy as long as the kinetic term cannot be neglected which, however, pushes these states out of the Fock space, analogous to the phenomenon associated with the second confluence process described in section \ref{sec-limit}. 


\section{Conclusions}\label{sec-concl}
We have re-examined the Rabi-Stark model using a scale transformation to an effective quantum Rabi model. The resulting $G$-function has a pole structure indicating for $\g\ra\om_-$ the emergence of a ``big" continuum above the threshold energy $E_c=-\D-g^2/\om$, extending  to positive infinity on the real energy axis. Previous studies of the isolated point $\g=\om$ \cite{maciejewski2015,xie2019,li2023} do not yield this continuum but instead
a set of discrete energy levels separated by a ``small" continuum for energies $-\D > E >E_{thr}=-\D-2g^2/\om$. We have shown analytically that this result is associated to a confluence process of the Schr\"odinger equation which postulates a bounded photon content for all eigenstates. Allowing for unbounded photon content even for states with low energy leads to another confluence process and a concomitant representation of the canonical commutation relations (\ref{alta}) which is not unitarily equivalent to the standard one in the limit $\g\ra\om_-$, a phenomenon well-known in quantum field theory \cite{segal1970}. This alternative confluence process leads to the ``big" continuum, as suggested by the $G$-function, but with the lower threshold energy $E_{thr}=-\D-2g^2/\om$. Because the first confluence is \emph{still valid}, we obtain from it not only the ``small" as part of the ``big" continuum, but also two sets of discrete states, the LBS which are true bound states below $E_{thr}$ but exist only in a limited parameter regime, and the BICs above $E=-\D$ which are embedded into the ``big" continuum. This situation bears a remarkable resemblance to the original atomic Stark effect describing the hydrogen atom in a static and spatially constant electric field which features also ``bound states"  which hybridize with the surrounding continuum and are thus only metastable \cite{hehenberger1974,cerjan1978,glushkov1993}.

The results obtained via the two confluence prosesses are corroborated by exact diagonalization of the Hamiltonian in state spaces with large dimension. For values of $\g/\om$ up to $0.999$, the spectral graphs exhibit a preContinuum of very dense levels together with a discrete set of states showing narrow avoided crossings with the preContinuum: the preBICs. The preBIC energies are well approximated by the BIC energies already for $\g=0.9$ and $0.99$, while the lower bound states which occur only for small $\Delta$ have precursors only for $\g$ much closer to $\om$. The threshold energy $E_{thr}$ is approximately reached for $\D=0.05$ and $\g=0.999$.

We have also investigated the model with a different analytical approximation based on effective potentials  which should be valid for large $\D/\om$ but turns out to give excellent qualitative insight for arbitrary parameters. Especially, it is possible to identify the preBICs with the eigenstates of the upper potential surface and the preContinuum with eigenstates of the lower one. The threshold energy for the single continuum appearing here in the limit $\g\ra\om_-$, comprising both the ``big" and the ``small" from section \ref{sec-limit}, is the same as obtained by the other methods.  

The QRSM is well-defined for all parameter values, including the regime $|\g|>\om$, because the Hamilton operator is always self-adjoint, although it has no ground state for $|\g|>\om$, similar to the two-photon QRM \cite{braak2023}. Therefore it can be studied  without the regularization by an additional confining potential proposed in \cite{li2023}. 
The central feature of the vicinity of the critical point, energetically low-lying states with very high photon content hybridizing with states having low photon content, deserves further study both theoretically and experimentally, the latter in view of possible applications for quantum technologies. Exact diagonalization of the QRSM using state spaces with high truncation number should allow to study the preContinuum and the gradual transition from normalizable states in $l^2(\Rr)$ to states outside of the Hilbert space, forming eventually the spectral continuum at the critical point. 

 \begin{acknowledgments}
D.B. acknowledges funding by the Deutsche Forschungsgemeinschaft (DFG, German Research Foundation) under grant no. 439943572. Computations were carried out on the Vera PC-cluster at Chalmers Center for Computational Science and Engineering with support from the Department of Physics at the University of Gothenburg.
\end{acknowledgments}
\vspace{4mm}\par
 $^{\ast}$ Corresponding author. Email: daniel.braak@uni-a.de
 
$^{\ast\ast}$ Corresponding author. Email: congllzu@gmail.com

 

\end{document}